\documentclass[apj]{emulateapj}

\slugcomment{}

\shorttitle{}
\shortauthors{}

\usepackage{color}
\usepackage{amsmath}
\usepackage{enumitem}

\begin{document}
\textheight 23.75cm

\title{Populations of Young Stellar Objects in Nearby Molecular Clouds}

\author{Tien-Hao Hsieh and Shih-Ping Lai}
\affil{Institute of Astronomy, National Tsing-Hua University, Hsinchu 30013, Taiwan\\
shawinchone@gmail.com\\
slai@phys.nthu.edu.tw}

\begin{abstract}
We develop a new method to identify YSOs from star-forming regions using the photometry data from Spitzer's c2d Legacy Project. The aim is to obtain YSO lists as complete as possible for studying the statistical properties, such as Star Formation Rate (SFR) and lifetimes of YSOs in different evolutionary stages. The largest obstacle for identifying YSOs comes from background galaxies with similar SEDs to YSOs. Traditionally, selected color-color and color-magnitude criteria are used to separate YSOs and galaxies. However, since there is no obvious boundary between YSOs and galaxies in Color-Color Diagrams (CCDs) and Color-Magnitude Diagrams (CMDs), those criteria may exclude faint YSOs near the boundary. In this paper, we separate the YSOs and galaxies in multi-dimensional (Multi-D) magnitude space, which is equivalent to using all variations of CMDs simultaneously. Comparing sources from molecular clouds to Spitzer's SWIRE data, which have negligible amount of YSOs, we can naturally identify YSO candidates locating outside of the galaxy populated regions in the Multi-D space. In the five c2d-surveyed clouds, we select 322 new YSO candidates (YSOc), miss/exclude 33 YSOc compared to \citet{ev09} and result in 1313 YSOc in total.   As a result, SFR increases 28\% correspondingly, but the lifetimes of YSOs in different evolutionary stages remain unchanged.  Comparing to theories \citep{kr05}, our derived SFR suggests that star formation in large scale is dominated by supersonic turbulence rather than magnetic fields.  
Furthermore, we identify 7 new Very Low Luminosity Objects.
\end{abstract}



\section{INTRODUCTION}
A full census of the Young Stellar Object (YSO) population in star-forming regions is essential for accurately calculating statistical properties, such as Star Formation Rate (SFR) and lifetimes of YSOs in different evolutionary stages, which are fundamental parameters for assessing the physical mechanism of global star formation.
The idea that SFR should be related to gas surface density was first proposed by \citet{sc58} and the relation is measured in a galaxy sample by \citet{ke98}, which is known as Kennicut--Schmidt relation. To examine this relation in our Galaxy, \citet{ev09} count the numbers of YSOs in Giant Molecular Clouds (GMCs), and assume a mean mass and formation timescale for these YSOs. 
They find that their measured SFRs are higher than that indicated by the Kennicut--Schmidt relation, and claim that is because the Kennicut--Schmidt relation applies to average over much larger regions than individual clouds. \citet{he10} extend the sample and obtain a similar result to that from \citet{ev09}. In addition, \citet{la10} find that the SFR is linearly proportional to the mass of cloud above an extinction threshold and this relation shows excellent agreement between galactic and extragalactic star-forming activity.

However, what physical mechanism determines the SFR is still unclear.
It has long been known that the observed SFR is too low for a GMC collapsing on its free-fall timescale without support against gravity \citep{zu74}. 
Supporting force such as supersonic turbulence and/or magnetic fields has been used to explain the low SFR;
however, observational data have not been able to determine which is the dominated supporting force.
\citet{kr05} and \citet{kr07} derive analytical relations between SFR and viral parameter
(the ratio of kinetic energy and gravitational energy), and show that SFR is linearly proportional 
to viral parameter for magnetic field dominated scenario while SFR decreases exponentially
versus viral parameter for turbulence dominated scenario. 
Here we attempt to obtain SFR as accurate as possible by developing 
a new YSO census method.   In addition, a better census method will provide 
better estimates in the lifetimes of YSOs in different evolutionary stages, 
which are usually estimated by comparing the fractions of YSOs in each stage, 
and successful theories should be able to explain the observed YSO lifetime scales.

Spitzer Space Telescope \citep{we04} provides high sensitivity surveys with the Infrared Array Camera (IRAC) at 3.6, 4.5, 5.8, and 8.0 $\mu$m \citep{fa04} and the Multiband Imaging Photometer (MIPS) at 24, 70, and 160 $\mu$m \citep{ri04}, which allow us to search for YSOs in star-forming regions.
Identifying YSOs is usually achieved by removing stars and background galaxies in the data \citep{yo05,ha07,re07,re10}.
Stars can be easily selected by fitting SEDs with reddened stellar atmosphere.
However, the SED morphologies of galaxies are very similar to those of YSOs at infrared wavelengths \citep{ha06}, 
and Spitzer can detect substantial amount of background galaxies because of its high sensitivity.
Therefore, separating background galaxies and YSOs becomes a difficult problem.
Several methods have been developed to identify YSOs from the molecular clouds \citep{ha07,gu05,re10} using Spitzer data.
These methods identify YSOs by eliminating the background galaxies based on comparing the distributions of the observed data with galaxy data in Color-Color Diagrams (CCDs) and Color-Magnitude Diagrams (CMDs).
However, there are no obvious boundaries between YSOs and galaxies in CCDs and CMDs.
Hence, based on different considerations, different works use different sets of CCDs and CMDs, and set their own boundaries in CCDs and CMDs to identify YSOs.
Therefore, sources located close to the boundaries are possibly classified as different objects using different methods.

A large YSO survey toward five nearby molecular clouds has been done by the Spitzer Legacy Project ``From Molecular Cores to Planet Forming Disk'' (c2d; Evans et al. 2003).
The c2d project used an unnormalized galaxy probability \citep{ha07} to eliminate possible galaxies and identify YSO candidates (YSOc) from the five clouds. (In the paper, YSOc is used for sources selected by our method or the c2d project, which may or may not be confirmed as true YSOs by other studies.)
This galaxy probability is primarily calculated from the locations of sources in three CMDs by comparing to the Spitzer Wide-Area Infrared Extragalactic Survey (SWIRE) data \citep{lo03}, which contain negligible amount of YSOs because the observations were towards high galactic latitude.
Since this galaxy probability was calculated only for sources detected in all IRAC and MIPS1 (24 $\mu$m), the c2d YSO catalog may miss faint
YSOs which are undetected at one or more of the five bands.
In this paper, we develop a new YSO identification method and apply it to the five c2d-surveyed molecular clouds, providing a more complete YSO catalog in these clouds.

With a relatively complete YSO sample, we will be able to identify more Very Low Luminosity Objects (VeLLOs), which are defined as protostars with internal luminosity, L$_\textmd{int}$, smaller than 0.1 L$_\odot$.  The first VeLLO, L1014-IRS, was observed by \citet{yo04}. \citet{du08} then design a set of color criteria and also use the c2d galaxy probability to select faint Class 0 and early Class I sources as VeLLO candidates.  The faint nature of VeLLOs suggests that they could be either very young protostars or very low mass protostars, but recent works have shown that the low luminosity can be explained by  protostars in the quiescent phase of episodic accretion processes \citep{du10}.
To identify more VeLLOs, \citet{du08} use the galaxy probability from c2d project but adopt a higher cutoff, which allows to find more faint sources; the galaxy probability are calculated based on the general fact that the galaxies are faint \citep{du08,ha07}. However, the change of cutoff may result in higher galaxy contamination and more analysis are essential for confirming those VeLLOs of \citet{du08}. 
Therefore, a method that identifies faint YSOc naturally may help to reduce the galaxy contamination.

We develop a new method, multi-dimensional (Multi-D) method to identify YSOs naturally.
While a CMD (magA, magB-magC) contains only information from the three consisting magnitudes, Multi-D magnitude space contains all the information of the consisting magnitudes.
We treat the number distribution of our galaxy sample in Multi-D magnitude space as galaxy probability distribution and sources located in the region without galaxies are classified as YSOc.
We apply this method to the five c2d-surveyed clouds and the number of YSOc sample is increased by a factor of 28\%.
Therefore, the SFR we calculated in this work are 28\% larger than that calculated by \citet{ev09} due to the increased number of YSOc,
but the life times of YSOs in different evolutionary stages remain unchanged due to the similar ratios between different stages of newly found YSOc.
We describe the data we used in \S 2 and our YSO identification method in \S3. In \S 4, we discuss how reliable our new YSOc are. The statistical analysis and discussion are presented in \S 5 and the conclusions are summarized in \S 6.

\section{DATA}

The data we used in this paper are from two Spitzer Legacy Projects, c2d and SWIRE.
c2d project observed five nearby molecular clouds and SWIRE project observed extragalactic fields that contain negligible amount of YSOs.
Our aim is to identify YSOs from the c2d data by comparing them to SWIRE data in Multi-D magnitude space.

\subsection{c2d Data}
Perseus, Serpens, Ophiuchus, Lupus (I, III and IV) and Chamaeleon II have been observed by c2d project in IRAC1--4 (3.6, 4.5, 5.8 and 8.0 $\mu$m) and MIPS1--3 (24, 70 and 160 $\mu$m) bands.
Hereafter, we use IR and MP to represent IRAC and MIPS, e.g., IR1 as IRAC1.
All data are processed with the c2d standard pipeline and these processes are described in the c2d data delivery document in detail \citep{ev07}. The pipeline extracts point sources from images at all bands and performs photometric measurements.
The Spitzer photometry data are merged with 2 Micron All-Sky Survey (2MASS) data (J, H, K bands) into source catalogs using a 2\arcsec~matching radius.
These source catalogs and images are released to Spitzer Heritage Archive.
Detail studies of individual clouds using these catalogs have been presented in several papers (e.g., Perseus: J\o rgensen et al. 2006; Serpens: Harvey et al. 2007a; Ophiuchus: Padgett et al. 2008; Lupus: Mer\'{i}n et al. 2008; Chamaeleon II: Porras et al. 2007).
Two kinds of catalogs, FULL and HREL (High Reliability), are provided by the c2d project.
FULL catalogs include all detected sources but some of them could be fake detection misidentified by automated photometry processes.
Sources in HREL catalogs have to be detected in one band with S/N $>$ 7 accompanied by a detection in another band with S/N $>$ 5. In this paper, we select YSOc only from HREL catalogs in order to avoid fake sources.

We adopt the photometry results from the c2d catalogs for our analysis.
Source flux in the c2d catalogs are measured with Point Spread Function (PSF) fitting.
An important factor, image type (imtype), is used to indicate the quality of PSF fitting.
Sources can be well fitted with PSF are assigned imtype = 1 and sources with imtype $\neq$ 1 have high uncertainty in photometry resulting from low S/N ratio, extended size, or non-circular shape, etc \citep{ev07}.
For example, imtype = 7 indicates that the S/N ratio is too low to be tested for shape and 
we found that it is mostly due to the confusion from surrounding cloud structure.
Therefore, we will check the images of sources with imtype $\neq$1 to determine whether they are real sources (see \S 3). \\
\subsection{SWIRE Data}
The SWIRE data used in this work are the same data used by \citet{ha07}, which
was processed with c2d pipeline to eliminate the systematic uncertainty introduced 
from different data reduction processes. 
Therefore, it can be used to compare with the c2d data fairly.
The SWIRE data used here are the observation toward ELAIS N1 extragalactic field which is at high latitudes (b $\sim$ 44$^{\circ}$) and the observed field has coverage of 5.3 deg$^2$ using both IRAC and MIPS.
Since the YSOs are concentrated in the galactic plane, there should be almost no YSOs in the SWIRE region.
In order to acquire a galaxy sample as pure as possible, we removed the stars in the SWIRE data using the same method to that for c2d data (\S 3.2).
Thus, it can be used to represent a full collection of the SED features of background galaxies.

\section{YSO IDENTIFICATION}

\begin{figure}
\includegraphics[scale=.33]{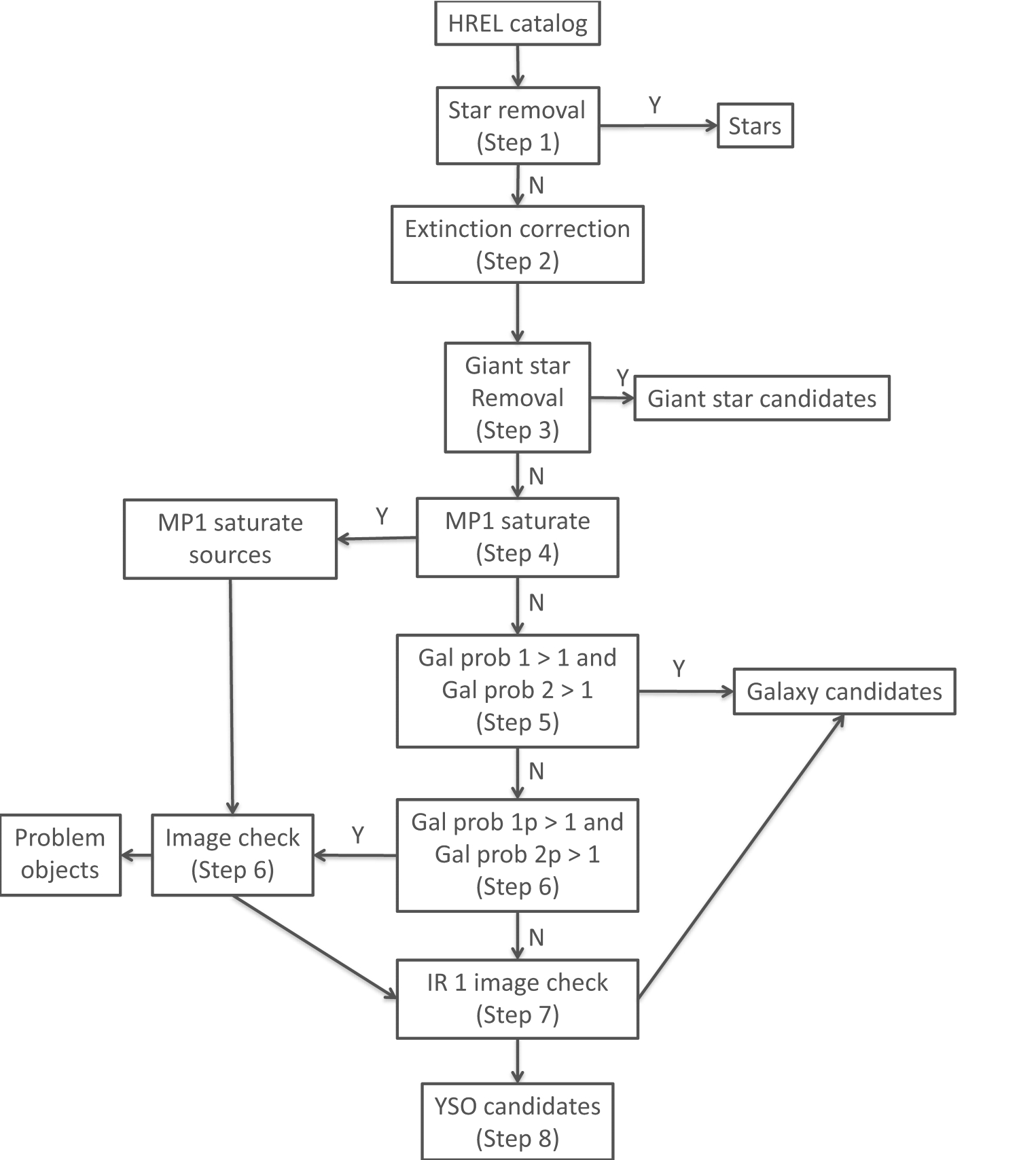}
\caption{The YSO identification process. Sources is consistent with the criteria in the block will enter the block connected by ``Y'' arrow and ``N'' for inconsistent.}
\end{figure}

\begin{figure}
\includegraphics[scale=.5]{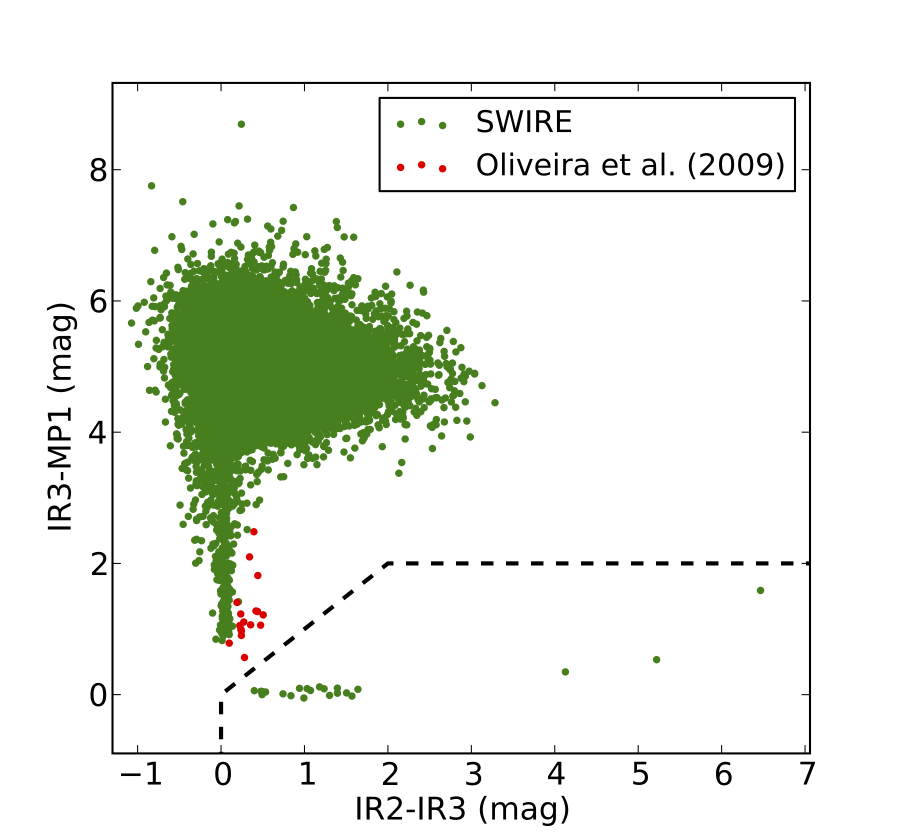}
\caption{The giant star selection criteria. Green points are the SWIRE data excluding stars and red points are giant stars in Serpens identified by \citet{ol09}. Sources located in the region below the dashed line are classified as giant stars.}
\end{figure}

\subsection{Selection Process}
Our YSO identification procedure is shown in Fig.\ 1 and described below. Basically, this process identifies YSOc by eliminating the non-YSO sources.\\
\begin{enumerate}[leftmargin=0.4cm]
\item  The main sequence stars were first removed by SED fitting. The fitting process is the same as that in \citet{ha07b} and \citet{ev07}, which select stars with reddened stellar atmosphere SED templates. 
The extinction law with R$_\textmd{V}$ = 5.5 from \citet{we01} is used to obtain the best fit extinction value, A$_\textmd{V}$. 
Weingartner and Draine R$_\textmd{V}$ = 5.5 model is suggested to be a good description for dusts in dense molecular clouds \citep{ch09}.
\item Since our galaxy sample (SWIRE) is observed toward regions with negligible extinctions, we de-reddened the whole c2d catalog in order to compare it with our galaxy sample.
We use the same computer program developed by the c2d project \citep{ev07} to construct the extinction maps from the extinction of background stars. The average errors of all pixels in each extinction map are 0.03--0.05 magnitude for each cloud.
Sources in the c2d catalog are then de-reddened according to the extinction value of its position in the map. We use the de-reddened fluxes in this paper except for the analysis with $\alpha$ (section 5.3.2).
\item Giant stars were removed using a CCD, IR2 -- IR3 versus IR3 -- MP1 (Fig.\ 2), as described in \S3.3.
\item We found that some sources labeled as ``U" (undetected) for MP1 in the c2d catalog are in fact saturated. 
Since bright MP1 flux will make the source far away from the galaxy populated region in the Multi-D array, we believe these sources are very young YSOs rather than galaxies.   Therefore, we classify these MP1 saturated sources as YSOc.
To find the saturated sources, we search all sources labeled as ``U'' at MP1 with pixel value larger than 800 MJy sr$^{-1}$ within 7.66\arcsec radius.  The cutoff 800 MJy sr$^{-1}$ is determined from experience by checking the pixel values around many saturated sources, and we found that it is small enough for us to pick up all saturated sources and also not too small for us to obtain too many fake sources.  The 7.66\arcsec ~radius is the 3$\sigma$ radius of MP1 PSF. 
Then, we examine the images of selected sources by eyes to confirm whether they are indeed saturated sources and the excess bright flux is indeed from a saturated source but not from a bad pixel.
The typical images of saturated sources are PSF-like with saturated holes (no values or low values) in centers and bright rings around the centers.
\item To eliminate background galaxies, many Multi-D arrays are constructed for the SWIRE data to calculate the smoothed galaxy density. According to the location of the c2d sources in the Multi-D arrays, we calculate four galaxy probabilities: Gal prob 1, Gal prob 2, Gal prob 1p and Gal prob 2p, where Gal prob 1 and Gal prob 2 are the galaxy density in (J, Ks, IR2, IR4 and MP1) and (IR1, IR2, IR3, IR4 and MP1) arrays or their subarrays (see \S3.2 for detail), and Gal prob 1p and Gal prob 2p are the galaxy density from the arrays discounting the bands with imtype $\neq$ 1. Sources with both Gal prob 1 and 2 $\geq$ 1 are classified as galaxy candidates and removed.
\item The remaining sources are with Gal prob 1 $<$ 1 or Gal prob 2 $<$ 1, thus are potential YSOc.
Images of sources with Gal prob 1 (2) $<$ 1 but Gal prob 1p (2p) $\geq$ 1 (meaning an unreliable detection would affect its galay probability) will be examined by eyes. Sources with unreliable photometry such as jet knots or cloud structures are removed (see \S 3.2).
\item For all the remaining sources, we also check the IR1 images to eliminate the nearby resolved galaxies which can be bright and thus with low galaxy probability. Eleven nearby galaxies are removed through this process.
\item The remaining sources are classified as YSOc.
\end{enumerate}

\subsection{Multi-D Array Construction and Galaxy Probability Calculation}
The heart and soul of our new method is to use the galaxy density in the Multi-D space as an indicator of the ``galaxy probability''; anything outside of the galaxy populated region would have galaxy density smaller than 1 and thus will be selected as YSOc.  We assume that SWIRE data is large enough to contain a complete set of galaxy sample.  Note that our ``galaxy probability'' here is not a real probability for a source to be a galaxy, it is in fact an unnormalized number which indicates how many galaxies near a specific position in Multi-D space.  In this section, we describe the details for constructing Multi-D space and defining the galaxy probabilities. 

The data we used contain ten bands (J, H, Ks, IR1, IR2, IR3, IR4, MP1, MP2 and MP3).
Although a high-dimensional array includes more information than that of a low-dimensional array,
the maximum dimension of the data array we can handle is limited by our computing resource.
If we want to calculate the galaxy probability from all 10-band data using a 10-D array, 
our computer will need random access memory of $\sim$10$^8$ G byte (with the cell sizes we choose, see below), which largely exceed the limitation of our computer.
Therefore, we choose to analyze our data with two 5-D main arrays constructing from the bands that can represent the main SED features; one contains J, Ks, IR2, IR4 and MP1 bands (corresponding to Gal prob 1) and the other contains all IR1--4 and MP1 bands (corresponding to Gal prob 2).
MP2 and MP3 bands are excluded due to their low detection numbers and their poor angular resolutions (18\arcsec~and 40\arcsec, respectively).  
The J--MP1 array covers the widest range of wavelengths and the IR1--MP1 array is designed for selecting YSOc that have no 2MASS detections. The IR1--MP1 array is necessary for identifying young embedded YSOc since they are very faint at short wavelengths and are often undetected in 2MASS bands.

Because the galaxy locations are discrete in the Multi-D arrays, we need to carefully choose the cell size and also employ a smoothing process to produce a smoothed galaxy distribution, so that the galaxies fall in between SWIRE sources will not be classified as YSOc.
We find that using cell size of 0.2 mag for all bands is adequate for our data.
We smooth the data using Multi-D Gaussian beams with a peak value of 1, a standard deviation $\sigma$ = 2 cells (0.4 magnitude) in every dimension and an outer cutoff at radius of 7 cells.
The galaxy probability for a source located at (x$_1$, x$_2$, x$_3$, x$_4$, x$_5$) is defined as
the total number of galaxies, A(x$_1$, x$_2$, x$_3$, x$_4$, x$_5$), after smoothed with Multi-D Gaussian beams,

{\scriptsize
\begin{align}
&A_{smooth}(x_1, x_2, x_3, x_4, x_5)=   \nonumber \\ 
&~~~\sum_{x'_1=x_1-7}^{x'_1=x_1+7}~~\sum_{x'_2=x_2-7}^{x'_2=x_2-7}~~\sum_{x'_3=x_3-7}^{x'_3=x_3+7}~~\sum_{x'_4=x_4-7}^{x'_4=x_4+7}~~    \sum_{x'_5=x_5-7}^{x'_5=x_5+7} \nonumber \\
&~~~A(x'_1, x'_2, x'_3, x'_4, x'_5)e^{((\frac{x_1-x'_1}{2\sigma})^2+(\frac{x_2-x'_2}{2\sigma})^2+(\frac{x_3-x'_3}{2\sigma})^2+(\frac{x_4-x'_4}   {2\sigma})^2+(\frac{x_5-x'_5}{2\sigma})^2)},  \nonumber \\
&(x_1-x'_1)^2+(x_2-x'_2)^2+(x_3-x'_3)^2+(x_4-x'_4)^2+(x_5-x'_5)^2\leq49.\nonumber \\
\end{align}}
Therefore, Gal prob 1 = A$_{smooth}$ (J, Ks, IR2, IR4 and MP1) and Gal prob 2 = A$_{smooth}$ (IR1, IR2, IR3, IR4 and MP1).  If a galaxy is surrounded by other galaxies, the galaxy' galaxy probability should be larger than the peak value of smoothing beam, which is 1, because the smoothing process will accumulate the number of galaxies from nearby cells.
Therefore, the surface with galaxy probability = 1 is expected to enclose {the galaxy populated region} in Multi-D magnitude space except isolated galaxies.
For an isolated galaxy, its galaxy probability will be one. 
Out of 135400 SWIRE galaxies, there are only three galaxies with Gal prob 1 equal to one and no galaxies with Gal prob 2 equal to one, which indicates that the galaxy sample is well smoothed in the Multi-D arrays.  Therefore, sources located out of the smoothed galaxy distribution will have galaxy probability $<$ 1 and can be classified as YSOc.


Since a large number of sources are not detected in all five bands of these two 5-D arrays, for these sources, Gal prob 1 and Gal prob 2 are calculated from 3- and 4-D subarrays instead. 
Thus, sources with detections in three to five bands of the five bands for Gal prob 1 are assigned Gal prob 1 calculated from 3- to 5-D magnitude arrays, respectively, and same as Gal prob 2 (the detection threshold is set to S/N=2).
As a result, 15 subarrays (five 4-D and ten 3-D subarrays) for each 5-D array are constructed.
The array cell size and/or the smoothing lengths of the subarrays need to be modified, because the galaxy distribution is condensed from 5-D to 3- or 4-D and thus the number of galaxies in 3-D or 4-D array cell are much larger than that in 5-D array cell. However, it is very difficult to make the galaxy probability of each source be the same in different dimensions and fairly compare them in different dimensions. 
Therefore, we modified the smoothing beam to make the threshold of galaxy probability, 1, a reasonable number for separating YSOc and galaxies in arrays with different dimensions. 
Here we demonstrate the principle of beam size modification in the case of 2-D to 1-D.
We assume that there are very few galaxies close to the surface of galaxy populated region where the galaxy probability is $\sim$1.  
Considering two galaxies locating near the surface and along the diagonal direction, 
we find that the galaxy probability profile in 1-D is exactly the same as that in 2-D along the diagonal direction if we reduce the standard deviation of Gaussian beam in 2-D by a factor of $\sqrt{1/2}$ (Fig.\ 3).

\begin{figure}
\includegraphics[scale=.4]{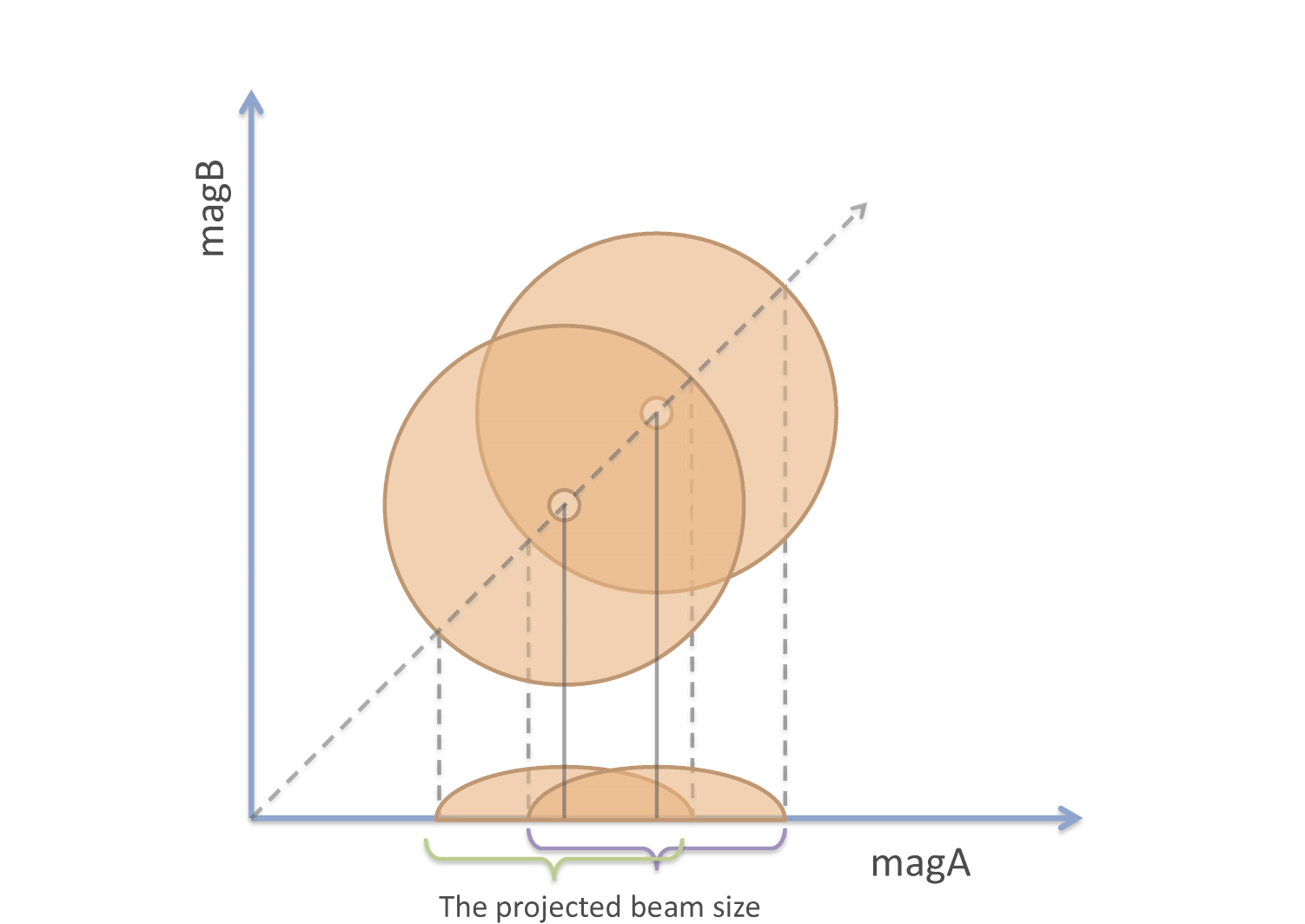}
\caption{
Reducing beam size from 2-D to 1-D. The two circles indicate the beam size in 2-D and the two half-ellipses indicate the projected beam size in 1-D. The ``galaxy probability'' along the dashed line, diagonal direction, in 2-D will be the same to that along x-axis in 1-D if a reducing beam length of $\sqrt{1/2}\sigma$ is applied.}
\end{figure}

Therefore, we use the standard deviation $\sigma$ of $\sqrt{3/5}$ $\times$ 2 cells (the $\sigma$ in 5-D array is 2 cells) for the 3-D arrays and $\sqrt{4/5}$ $\times$ 2 cells for the 4-D arrays, and we keep the cell size unchanged.
Such modification will produce similar galaxy probability distribution in the regions close to the boundary of galaxy populated regions for arrays with different dimensions, thus the same threshold of galaxy probability (i.e., 1) can be used to identify YSOc.

We found that a lot of sources with PSF fitting flag ``imtype $\neq$ 1'' (\S 2.1) have contaminated flux measurements due to jet knots, nearby bright sources and extended cloud structure, which may change the locations of the sources in the Multi-D arrays. 
Thus, we define Gal prob 1p and Gal prob 2p which are calculated from subarrays only using the detections with imtype = 1 (well-fitted as a point source), and use them along with Gal prob 1 and Gal prob 2 for separating YSOs and galaxies.
Since imtype=1 indicates that the source has a more precise flux measurement, source with Gal prob 1p (or 2p) $<$ 1 implies that this source locates out of the galaxy populated region in Multi-D space with high confidence. 
If a source has Gal prob 1 (or 2) $<$ 1 but Gal prob 1p (or 2p) $>$ 1, it is obvious that the bands with imtype $\neq$ 1 make the sources deviated from the galaxy populated region. For these sources, images of IR1--IR4 and MP1 bands are examined with eyes. 
If a reliable point source is found in the images at bands with imtype $\neq$ 1, the Gal prob 1 (or 2) calculated containing such bands is reliable and the sources are classified as a YSOc.
A large amount of examined sources were found to have unreliable photometry measurement (70\%--90\% for different clouds) due to jet knots, nearby bright sources and extended cloud structure, and are removed from our YSOc list.

\subsection{Giant Star Contamination}
Giant stars could be bright and have similar SEDs to Class II/III YSO in infrared \citep{ol09}, thus are very difficult to be removed only using IRAC and MIPS data.
We are not able to exclude the giant stars by finding the natural boundary of them in the Multi-D array, because the giant star sample is too small to represent a full collection of the SED features.
From SWIRE data, we found that there are 30 bright sources that are not group with the majority of the SWIRE sources
in the (IR2 -- IR3,  IR3 -- MP1) CCD.
Because we think that SWIRE data only contain galaxies, main-sequence stars, and giant stars, we suspect that these 30 bright sources are giant stars.  Therefore, we define 3 criteria to exclude the giant stars from the c2d data:
IR3 -- MP1 $<$ 2, IR2 -- IR3 $>$ IR3 -- MP1 and IR2 -- IR3 $>$ 0 (Fig.\ 2).
The numbers of giant stars selected from these criteria are 9, 8, 5, 2 and 1 in Serpens, Ophiuchus, Lupus, Perseus and Chamaeleon II, respectively.
However, these criteria do not remove 17 sources in Serpens which have been identified to be giant stars by \citet{ol09} 
based on the optical spectrum. Because the locations of these 17 sources in Fig.\ 2 are closer to the SWIRE sources than our selected giant stars, we suspect that these spectroscopically identified giant stars may have small disk or dusty envelopes that increase their IR3 -- MP1 color. We later removed these 17 sources from our YSOc list.
Since we did not find any spectroscopic data for giant stars in Perseus, Ophiuchus, Lupus and Chamaeleon II, we are not able to remove spectroscopic identified giant stars in these clouds.
However, because Serpens molecular cloud is located in the direction closed to the Galactic plane while other clouds are not, Serpens has the most serious giant star contamination among these clouds.
Thus, the giant star contamination should be much smaller in the other four clouds.

\begin{figure*}
\includegraphics[scale=.43]{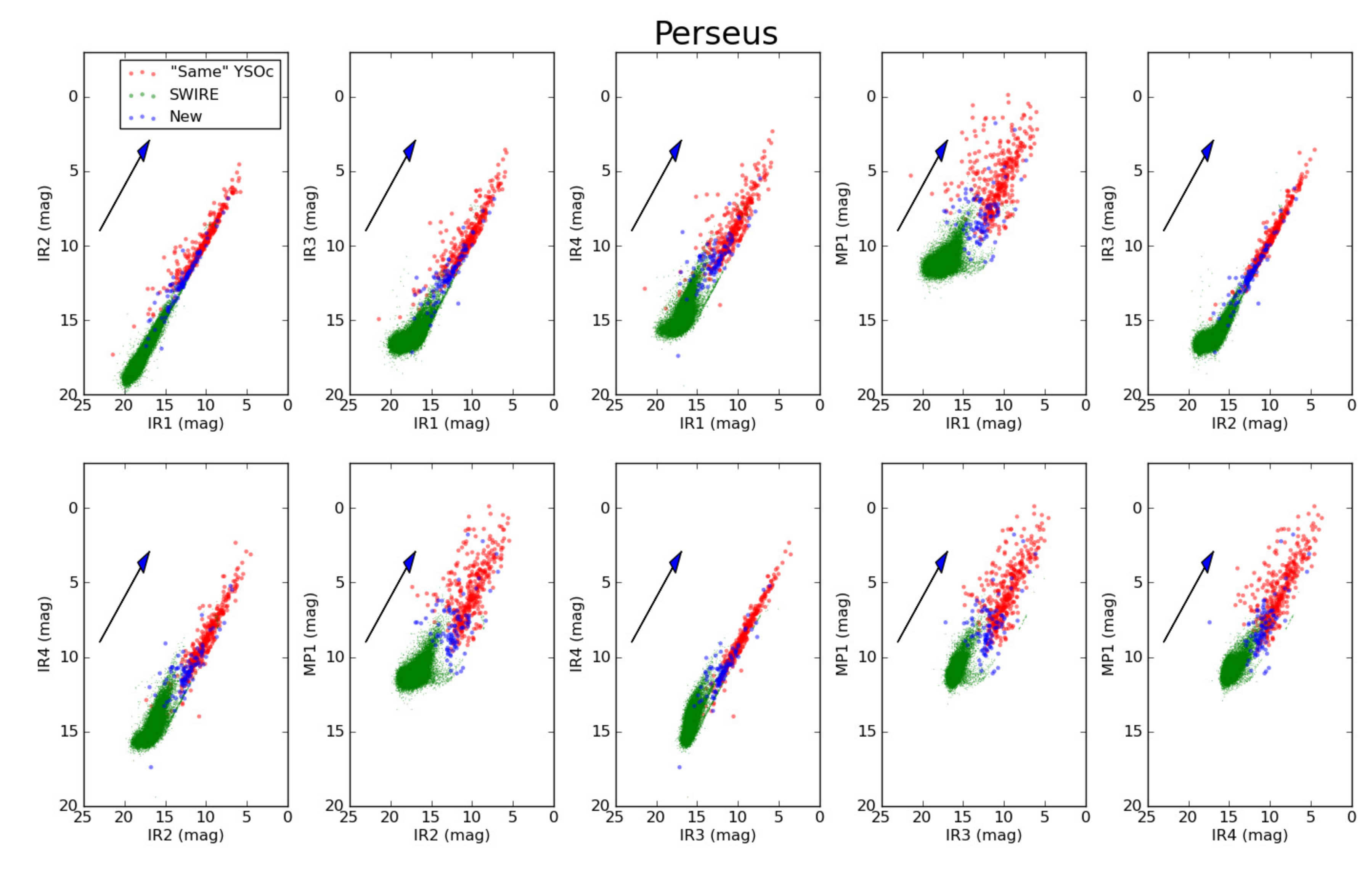}
\caption{
All variations of MMDs consisting any two bands of IR1, IR2, IR3, IR4, and MP1 for Perseus. Red, blue and green points indicate the ``Same'' YSOc, newly identified YSOc and SWIRE galaxies, respectively. The arrows represent the diagonal direction which is related to source brightness.}
\end{figure*}

\begin{figure*}
\includegraphics[scale=.43]{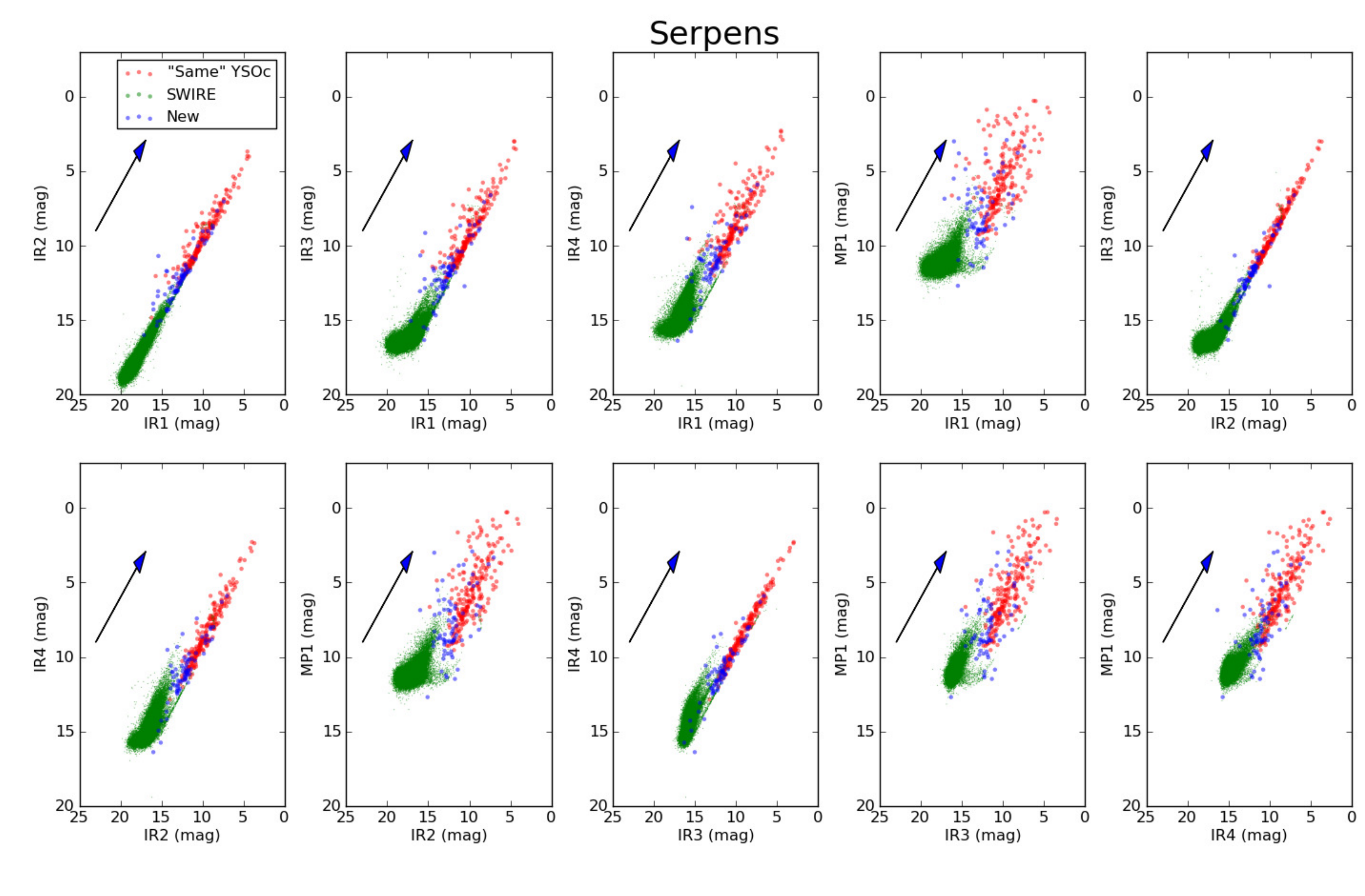}
\caption{The same to Fig. 4 for Serpens.}
\end{figure*}

\begin{figure*}
\includegraphics[scale=.43]{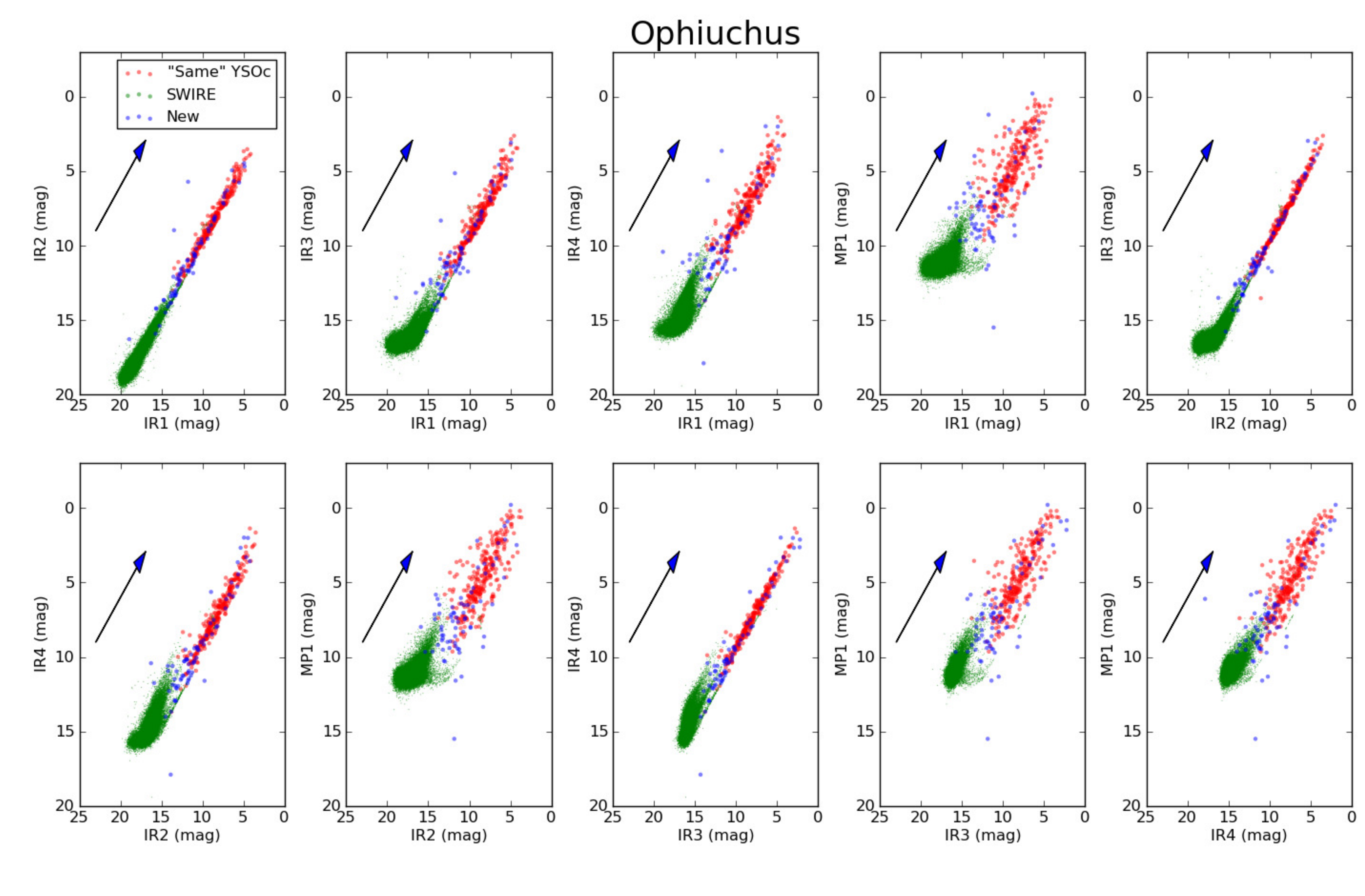}
\caption{The same to Fig. 4 for Ophiuchus.}
\end{figure*}

\begin{figure*}
\includegraphics[scale=.43]{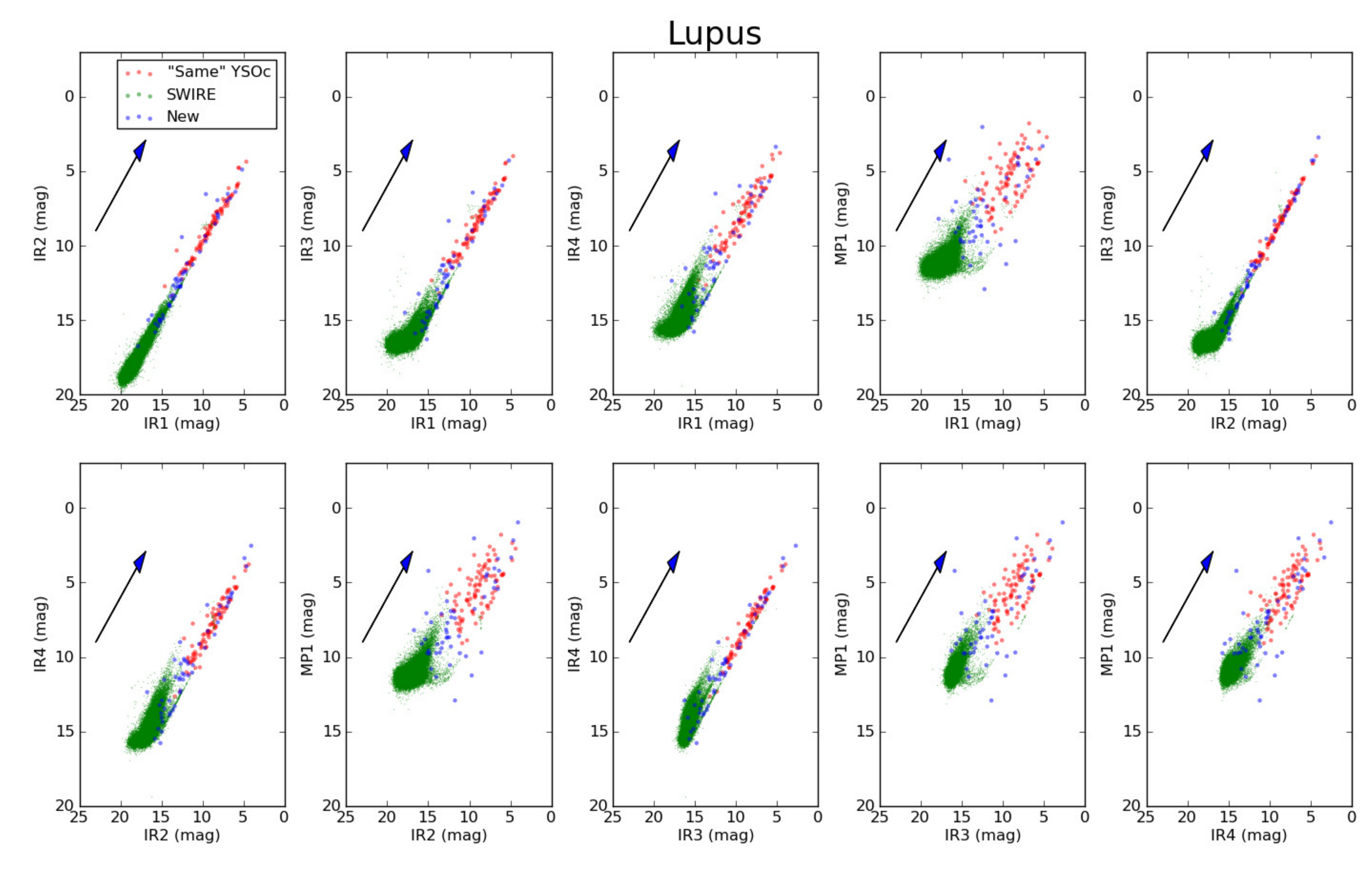}
\caption{The same to Fig. 4 for Lupus.}
\end{figure*}

\begin{figure*}
\includegraphics[scale=.43]{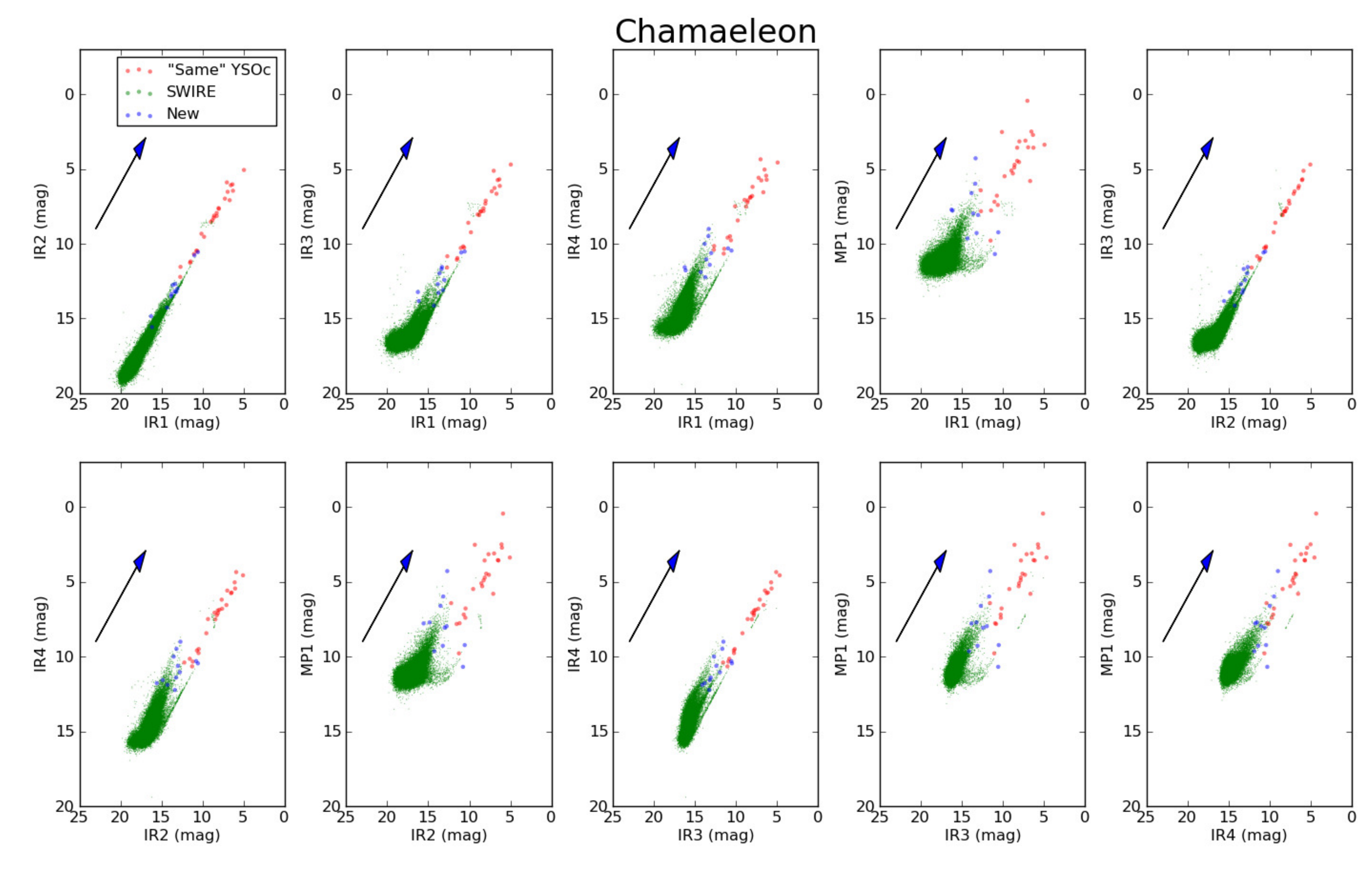}
\caption{The same to Fig. 4 for Chamaeleon II.}
\end{figure*}

\begin{deluxetable*}{p{2.0cm}p{2.5cm}p{2.5cm}p{2.5cm}p{2.5cm}p{2.5cm}}
\tabletypesize{\scriptsize}
\tablecaption{Numbers of selected YSOc and the comparison with the c2d lists}
\tablehead{
Cloud & YSOc selected with Multi-D Method (Group A) & YSOc in the c2d lists (Group B) & YSOc in both Group A and B (``Same YSOc'')\tablenotemark{a} & YSOc in Group A, but not in Group B\tablenotemark{a} & YSOc in Group B, but not in Group A}
\startdata
Perseus & 469 & 385 & 369 (78\%) & 100 (22\%) & 16 \\
Serpens & 296 & 227 & 218 (74\%) & 78 (26\%) & 9 \\
Ophiuchus & 367 &392 & 287 (78\%) & 80 (22\%)& 5 \\
Lupus & 143 & 94 & 91 (64\%) & 52 (36\%) & 3 \\
Chamaeleon II & 38 & 26 & 26 (68\%) & 12 (32\%) & 0 \\
Total & 1313 & 1024 & 991 (75\%) & 322 (25\%) & 33 
\enddata
\tablenotetext{a}{Percentage is calculated with respect to the YSOc number in Group A.}
\tablecomments{The YSOc number includes the three well-known sources that are not selected from Multi-D Method (Table 3).}
\end{deluxetable*}

\subsection{Advantages of Multi-D Method}
Using the Multi-D space as a whole instead of using limited numbers of selected CMDs to identify YSOs has the following advantages:
\begin{enumerate}[leftmargin=0.4cm]
\item Avoidance of selecting specific CMDs. In order to identify YSOs in molecular clouds, previous works adopted different CMDs as their selection criteria.
Although these CMDs are all selected with justifiable reasons, it is difficult to argue that which CMD separates YSOs and galaxies best and how many CMDs are sufficient for the identification.
The various selections of CMDs in use make the YSO identification results different by different works.
Since Multi-D magnitude space includes the information from all the possible variations of the CMDs, we have no need to select the specific CMDs. Therefore, Multi-D method is relatively complete and unbias compared to CMD methods.
\item A natural boundary of galaxy populated region in Multi-D space. 
We use the natural boundary constructed from number distribution of our galaxy sample.  This boundary encloses a relatively accurate galaxy SED ensemble, which reduces the bias in identifying YSOc near the boundary.
\item Uncovering YSOs that cannot be found with all possible CMDs. A CMD is equal or equivalent to a projection in Multi-D magnitude space. 
However, even all the possible variations of the CMDs are used, it is possible that some YSOs may not be found if they are located in a region without galaxies in Multi-D magnitude space but immerses in galaxy populated regions in all CMDs; that is, some YSOs cannot be revealed in any projections from Multi-D magnitude space (see Fig.\ 4--9).
\end{enumerate}
Therefore, Multi-D method provides the best opportunity to identify YSOs as complete as a photometric dataset can offer. 

\begin{figure}
\includegraphics[scale=.3]{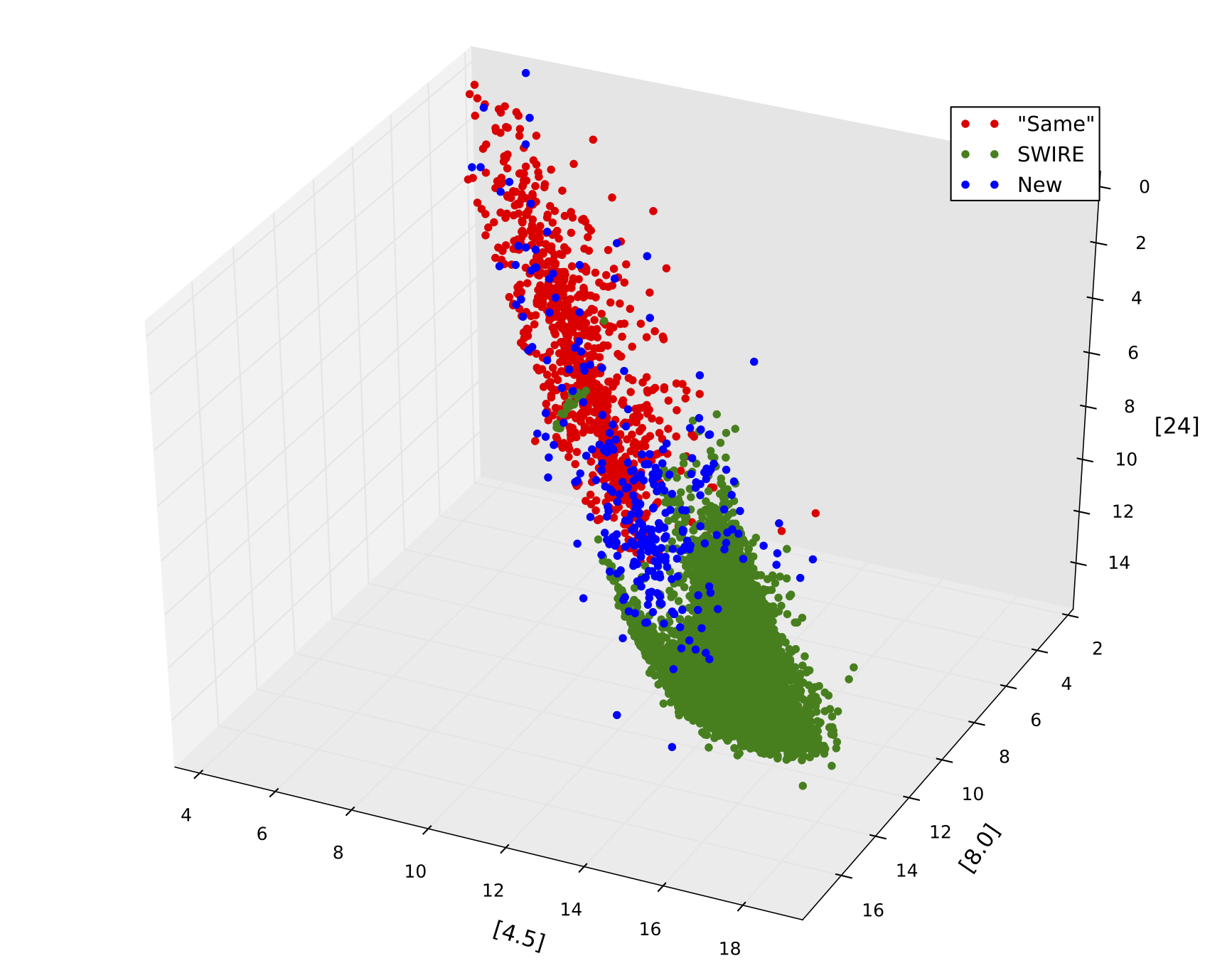}
\caption{The 3-D magnitude space with axes of IR2, IR4 and MP1. Color points indicate the sources as same as that in Fig. 4 and all YSOc in five clouds are plotted together.}

\includegraphics[scale=0.4]{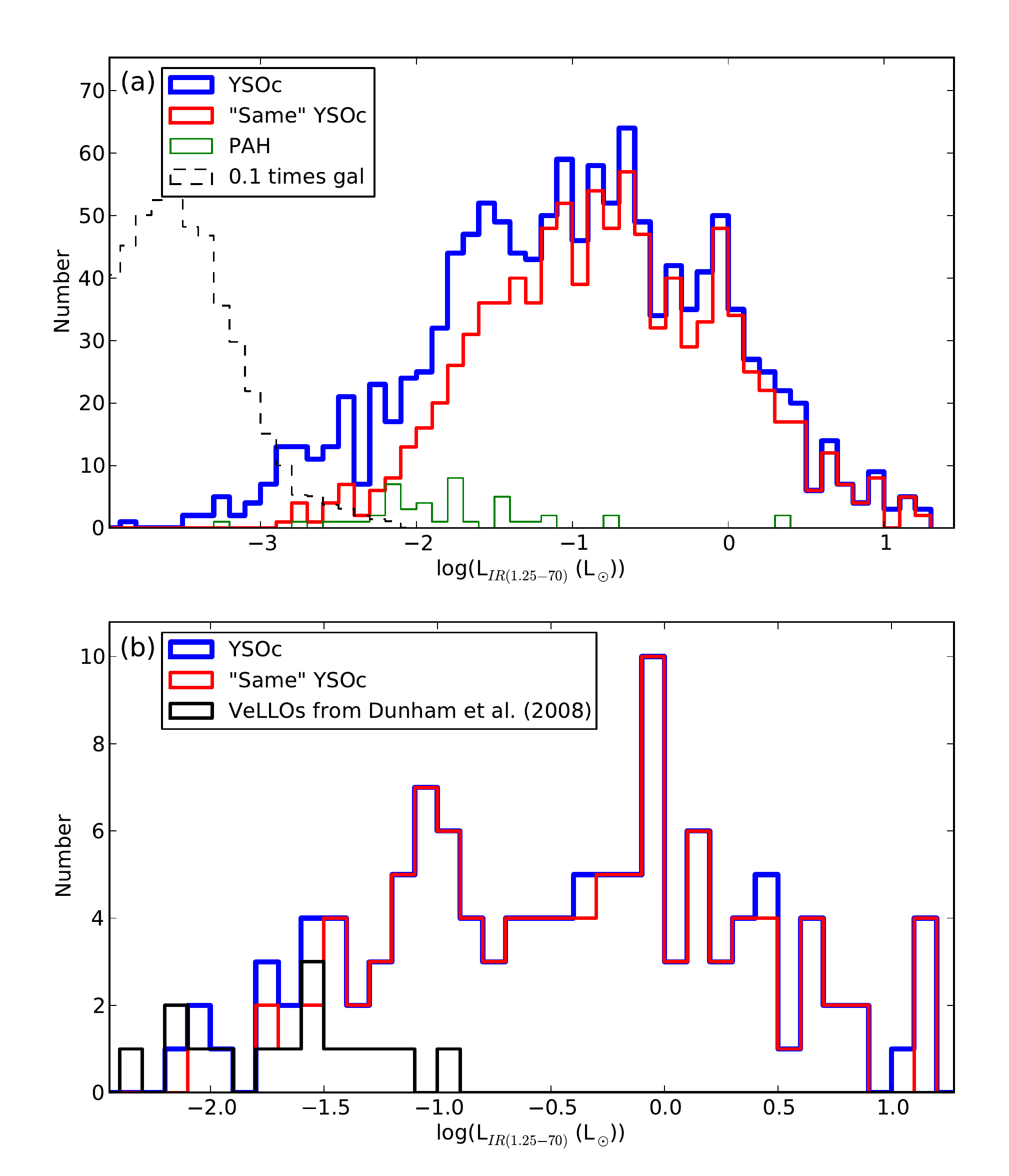}
\caption{(a) The distributions of infrared-luminosity for YSOc selected with Multi-D methods (blue), ``Same'' YSOc (red), YSOc with PAH feature (green), and background galaxies (black dashed line). The background galaxies are sources with both Gal prob 1 and Gal prob 2 $\leq$ 10 and their distribution is scaled by a factor of 0.1. (b) Same as (a), but only for Class 0 and Class I sources with MP2 detections and the black line are VeLLO candidates selected by \citet{du08} . 
}
\end{figure}

\section{RESULT: NEW YSOc LIST}
We present our YSOc lists for the five c2d-surveyed clouds selected with the Multi-D method.
The total number of our YSOc and the difference between our YSOc numbers and the c2d numbers are listed in Table 1, which shows that we increase the total YSOc number by 28\% (from 1024 to 1313). 
The YSOc we selected in Perseus, Serpens, Ophiuchus, Lupus, and Chamaeleon II, along with their galaxy probabilities are listed in Table 2. Note that the smaller a galaxy probability is, the more distant the sources is located away from the galaxy populated region.

\begin{deluxetable*}{ccccccccc}
\tabletypesize{\scriptsize}
\tablecaption{YSO candidates}
\tablehead{
\colhead{R.A.}           & \colhead{Dec}      & \colhead{Gal prob 1\tablenotemark{a}} &
\colhead{Gal prob 2\tablenotemark{a}}          & \colhead{Gal prob 1p\tablenotemark{a}}  &
\colhead{Gal prob 2p\tablenotemark{a}}          & \colhead{Ste den} &\colhead{YSO prob}  & \colhead{c2d classification\tablenotemark{b}}
\\
\colhead{degree}           & \colhead{degree}      & \colhead{log(num)} &
\colhead{log(num)}          & \colhead{log(num)}  &
\colhead{log(num)}          & \colhead{log(pc$^{-2}$)} &\colhead{log(num)}  & \colhead{}
}
\startdata
\multicolumn{9}{c}{Perseus}\\
\hline
51.1968071 & 30.4656086 & -1.99 & 2.44 & -1.99 & 2.44 & -0.06 & 1.03 & -- \\
51.3304656 & 30.7577636 & -- & -0.06 & -- & -0.36 & 0.47 & 0.46 & -- \\
51.3313375 & 30.5733784 & -1.36 & -$\infty$ & -- & -- & 0.43 & 0.95 & YSOc\_star+dust(IR1) \\
51.3430184 & 30.7538668 & -$\infty$ & -$\infty$ & -- & -- & 0.49 & 0.00 & YSOc\_red \\
51.4009155 & 30.7543631 & -$\infty$ & -$\infty$ & -- & -- & 0.44 & 0.23 & red \\
51.4020445 & 30.7561528 & -$\infty$ & -$\infty$ & -$\infty$ & -$\infty$ & 0.43 & 0.00 & YSOc\_red \\
51.4092048 & 30.4527241 & -0.46 & 1.14 & 0.38 & 1.86 & 0.04 & 0.01 & -- \\
51.4117940 & 30.7350532 & -$\infty$ & -$\infty$ & -- & -- & 0.51 & 0.03 & YSOc\_red \\
51.4130155 & 30.7328223 & -$\infty$ & -$\infty$ & -- & -$\infty$ & 0.52 & 0.01 & YSOc\_red \\
51.6561277 & 30.2578016 & -$\infty$ & -$\infty$ & -- & -- & 0.17 & 0.06 & YSOc\_red \\
51.8292253 & 30.2883997 & -0.32 & -0.44 & -- & -- & 0.63 & 0.14 & -- \\
51.8393548 & 30.8312253 & -2.22 & -$\infty$ & -- & -- & -0.16 & 0.00 & -- \\
51.9093741 & 30.2329495 & -$\infty$ & -$\infty$ & -- & -- & 0.87 & 0.87 & YSOc\_star+dust(IR4) \\
51.9109468 & 30.2267923 & -1.37 & -0.31 & -1.19 & 0.47 & 0.91 & 1.50 & -- \\
51.9117155 & 30.2235869 & -1.24 & -0.24 & -0.34 & 0.81 & 0.88 & 1.09 & -- \\
51.9128437 & 30.2175414 & -$\infty$ & -$\infty$ & -$\infty$ & -$\infty$ & 0.89 & 0.00 & YSOc\_red \\
51.9228069 & 30.3379885 & -$\infty$ & -$\infty$ & -- & -- & 0.43 & 1.64 & YSOc\_star+dust(IR2) \\
51.9301209 & 30.2080268 & -$\infty$ & -$\infty$ & -$\infty$ & -$\infty$ & 1.03 & 0.00 & YSOc\_red \\
51.9486420 & 30.2012577 & -$\infty$ & -$\infty$ & -$\infty$ & -- & 1.18 & 0.62 & YSOc\_star+dust(IR1) \\
52.0003793 & 30.1463963 & -$\infty$ & -$\infty$ & -- & -- & 0.87 & 1.36 & YSOc\_star+dust(IR2)
\enddata
\tablenotetext{a}{We calculate four galaxy probabilities: Gal prob 1, Gal prob 2,
          Gal prob 1p and Gal prob 2p, where Gal prob 1 and Gal prob 2 are the
          galaxy density in (J, Ks, IR2, IR4 and MP1) and (IR1, IR2, IR3, IR4
          and MP1) arrays or their subarrays (see \S3.2 for detail), and Gal
          prob 1p and Gal prob 2p are the galaxy density from the arrays
          discounting the bands with imtype $\neq$ 1.
}
\tablenotetext{b}{
If the c2d source type does not contains ``YSO'', the source is added to the YSOc list by \citet{ev09} based on ancillary data from literature. Sources without c2d classification, ``--'', are newly identified YSOc from this work.}
\end{deluxetable*}

We discuss the properties of our newly selected YSOc in \S 4.1 and argue that they are likely to be real YSOs.
In \S4.2, we present the reasons for excluding some YSOc from the c2d list and retrieve three missed established YSOs.   
In \S 4.3, we further discuss the YSOc with ``PAH-emission'' feature and label them as less reliable YSOc, which composes less than 3\% of all YSOc.
Finally in \S 4.4, we discuss the uncertainty resulting from our band selections for constructing two 5-D arrays.
\subsection{Analysis of The Newly Identified YSOc}
In order to examine whether our newly identified YSOc are likely to be real YSOs,
we compare our list and the c2d list in the following three aspects.  
First, we show the distributions of YSOc from the two lists are consistent in all MMDs and the two lists of YSOc have similar luminosity function. Second, we show that our YSOc stellar surface density are more consistent with the c2d's YSOc stellar density than galaxy surface density. Finally, we show that the distributions of the two YSOc lists are also consistent in the multi-dimentional color space. Note that these analyses can only give us the hints of that whether the newly identified YSOc are likely to be real YSOs or not. Without other data such as spectroscopy, we are not able to confirm whether they are real YSOs.
\subsubsection{Newly Identified YSOc in The Magnitude-Magnitude Diagrams and Their Infrared Luminosity Function}
Because we are not able to plot a Multi-D figure with more than 3 dimensions, we display the distribution of our newly identified YSOc in all possible MMDs,
in order to examine whether their distributions in the MMDs are similar to the previously identified YSOc. 
For the same type of sources with similar SEDs, their locations in the CCDs will be close to each other
and their locations in MMDs will be along the diagonal direction since the only difference between sources is the brightness (distance).   
Therefore, if the newly identified YSOc are real YSOs, they should align with the previously identified YSOc in the diagonal direction.
Fig. 4--8 show the populations of newly identified YSOc (blue), galaxy sample (green) and YSOc identified by both c2d and our method (red) (hereafter we call these YSOc as ``Same'' YSOc) in all variations of MMDs consisting any two bands of IR1, IR2, IR3, IR4, and MP1.
As we expected, the newly identified YSOc are well aligned with the ``Same'' YSOc in the diagonal direction, but at fainter ends close to the galaxy populated regions.
Although in Fig.\ 4--8, some faint YSOc are appeared to be mixed with galaxies, they are in fact separated in Multi-D space.
Fig.\ 9 shows an example of a 3-D space consisting of IR2, IR4 and MP1 magnitudes, in which YSOc and galaxies are better separated than in any 2-D space.
 
We also compare the infrared luminosity distribution of our YSOc list and the ``Same'' YSOc list (infrared luminosity is obtained from J to MP2), 
and find that most of the newly identified YSOc are at the faint end of the distribution (Fig. 10a). The distribution of the newly identified YSOc is more like an extension of YSOc sample at low luminosity end rather than part of the galaxy distribution.
Therefore, both the YSOc locations in MMDs and the luminosity distribution support that 
the Multi-D method are able to identify fainter YSOc which are not found by previous works.

\begin{figure*}
\includegraphics[scale=.73]{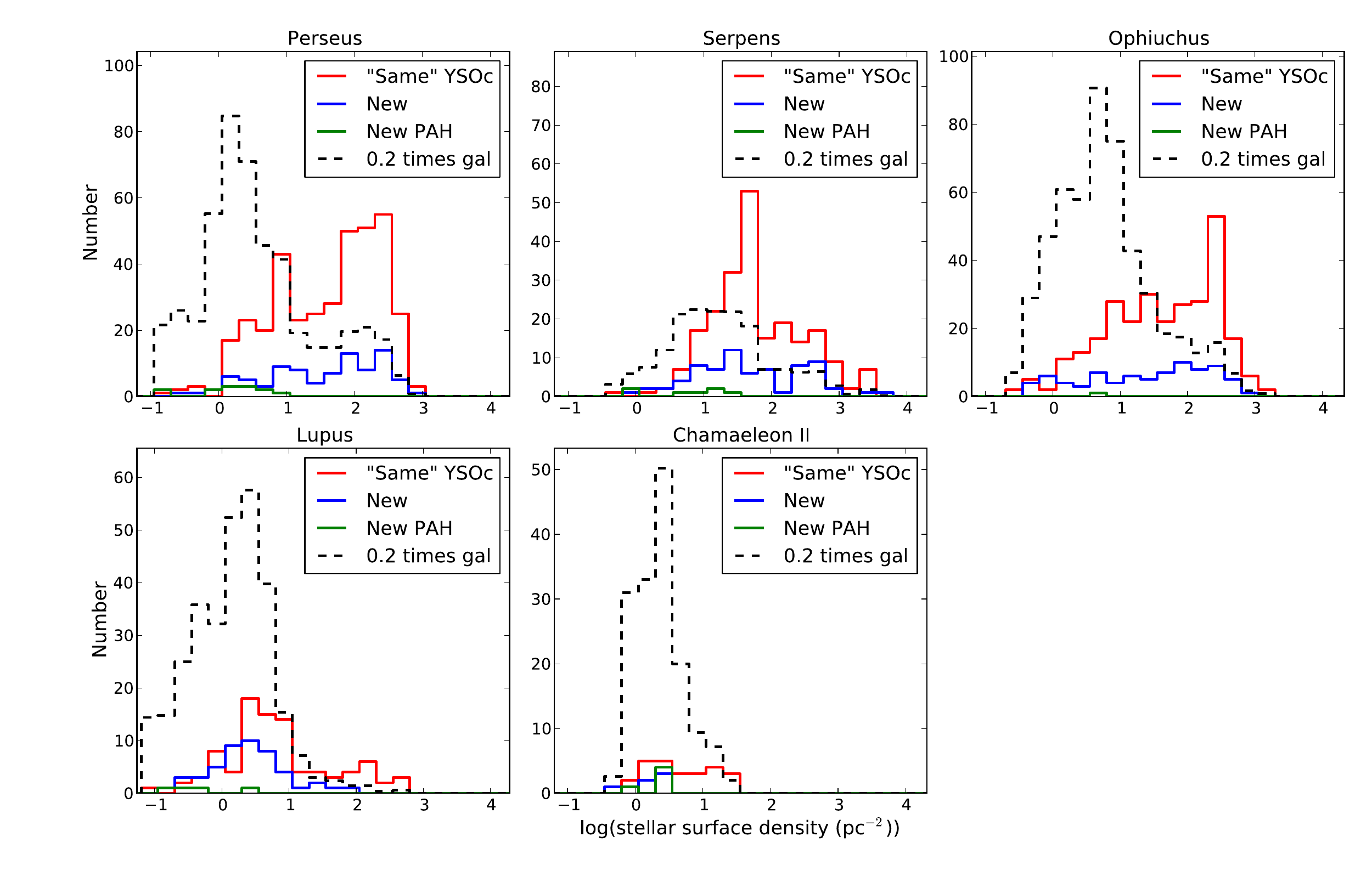}
\caption{The histograms of stellar (YSOc) surface density for sources in all clouds. The histograms indicate the ``Same'' YSOc (red), newly identified YSOc without PAH\_em feature (blue) and newly identified YSOc with PAH\_em YSOc (green), respectively. Black dashed line is the background galaxy found in the direction of the cloud and the number is multiplied by a factor of 0.2.}
\end{figure*}

\subsubsection{Stellar Surface Density}
Since YSOs tend to form in clusters in molecular clouds and background galaxies have relatively random distribution in the sky, the newly identified YSOc are expected to have higher probability to locate in the regions with other YSOc if they are real YSOs.   
We use stellar surface density as the parameter to indicate whether a newly identified YSOc are located within YSO clusters.  Stellar surface density is defined as 
\begin{equation}\label{}
\sigma\{\imath,\jmath\}=\frac{n-1}{\pi r_n^2\{\imath,\jmath\}}
\end{equation}
where n is the n-th closed star and $r_{n}$ is the distance to the n-th closed star \citep{gu05, gu09}.
Here we adopt n=6, same as that used in \citep{gu09}, as a surface density reference. 
The stellar surface density for each source is calculated from where they are with respect to the ``Same'' YSOc population, and the results are listed in column 7 of Table 2.  
In Fig. 11, we compare the stellar surface density for the ``Same'' YSOc sample, newly identified YSOc, and background galaxies for each cloud.  We  select background galaxies as sources with both Gal prob 1 and Gal prob 2 larger than 10.
We use Kolmogorov-Smirnov test to determine whether the stellar surface density distributions of our newly identified YSOc is similar to that of ``Same'' YSOc sample or galaxies. 

\begin{deluxetable*}{p{6cm}p{1.5cm}p{1.5cm}p{1.5cm}p{1.5cm}p{1.5cm}}
\tabletypesize{\scriptsize}
\tablecaption{K-S test results for stellar surface density - P value}
\tablehead{
& Perseus & Serpens & Ophiuchus & Lupus & Chamaeleon
}
\startdata
newly identified YSOc and ``Same'' YSOc & 0.88 & 0.08 & 0.21 & 0.01 & 0.07 \\
newly identified YSOc and galaxies & 0.00 & 0.00 & 0.00 & 0.06 & 0.37
\enddata
\end{deluxetable*}

The results (P-value) of Kolmogorov-Smirnov test are shown in Table 3. If the P-value is significantly larger than 0.05, we cannot reject the hypothesis that the distributions of the two samples are the same  (SciPy Reference Guide, 2012).
For Perseus, Serpens and Ophiuchus,
the P-values calculated from newly identified YSOc and ``Same'' YSOc are larger than that from newly identified YSOc and galaxies.
These results suggest that the stellar surface density distributions of our newly identified YSOc are like to that of ``Same" YSOc rather than background galaxies in Perseus, Serpens and Ophiuchus.
In Chamaeleon II and Lupus, the distributions of both newly identified and the ``Same'' YSOc sample show low stellar surface density;
therefore, the distributions have low statistical significance in determining 
whether newly identified YSOc are similar to real YSOs or background galaxies.

\begin{figure}
\includegraphics[scale=.56]{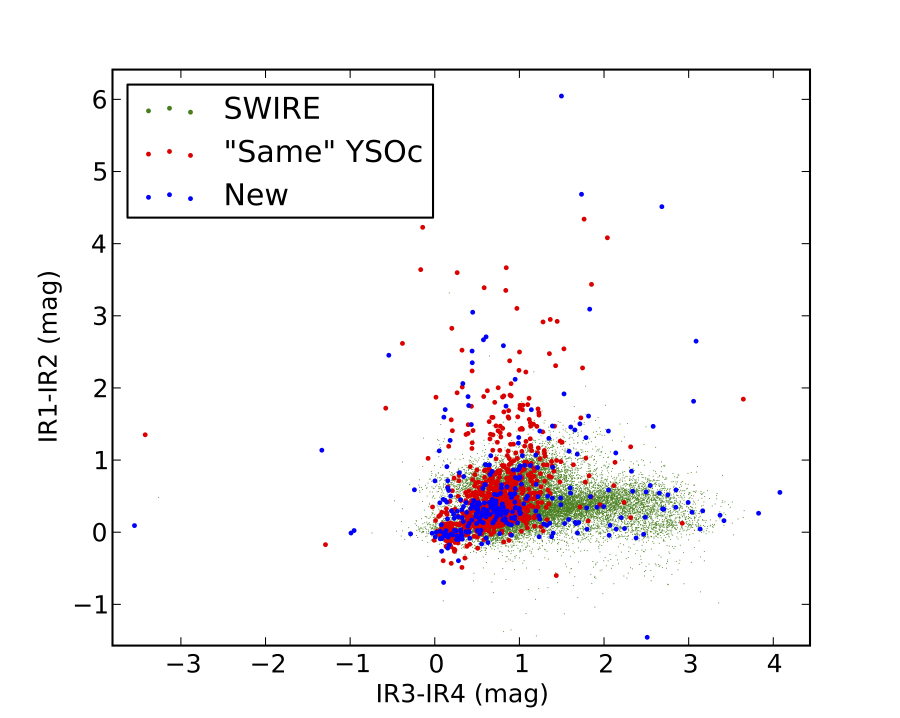}
\caption{Color-color diagram using IR1, IR2, IR3 and IR4 bands. Color points indicate the sources as same as that in Fig. 4 and all YSOc in five clouds are plotted together.}
\end{figure}

\begin{deluxetable*}{cc p{13cm}}
\tabletypesize{\scriptsize}
\tablecaption{True YSO missed by our selection method}
\tablehead{
\colhead{R.A.}           & \colhead{Dec}       &\colhead{reason description}
\\
\colhead{degree}           & \colhead{degree}      & \colhead{}
}
\startdata
193.3217719 & -77.1196340 &    DK Cha. This source saturates at IR2, IR4 and MP1 bands and is not a HREL source.  We consider only HREL sources (see \S 2.1). \\
246.6100815 & -24.4083265 &    VLA1623. Since it has only been detected at MP1 and longer wavelength, lack of data excludes it from our selection process. \\
248.0942748 & -24.4755023 &   IRAS16293-2422.   IRAS16293-2422A and B are not resolved at MP1.  IRAS16293-2422A is not detected in IRAC bands, while IRAS16293-2422B is barely detected from IR2 to IR4.    These sources are not HREL sources.
\enddata
\end{deluxetable*}

\subsubsection{YSO Probability}
If YSOs have similar SEDs, they should distribute continuously in a rather concentrated region 
in Multi-D color space (Fig.\ 12).
We examine this property by calculating a new parameter, YSO probability, in Multi-D color space, 
which is calculated in the way similar to galaxy probability in Multi-D magnitude space.
Here we calculate the YSO probability from the number distribution of our identified YSOc in Multi-D color space constructed with four axes (IR1-IR2, IR1-IR3, IR1-IR4 and IR1-MP1). 
Similar to the galaxy probability calculation, the subarrays (2- and 3-D arrays) are also used for YSOc with detections only in three and four of the all five bands. The cell size is 0.2 magnitude as same as that used in galaxy probability calculations, and the $\sigma$ of smoothing beam are 2, $\sqrt{3/4}$ $\times$ 2 and $\sqrt{2/4}$ $\times$ 2 cells for 4-D, 3-D and 2-D arrays, respectively. 
The YSO probability for each source is listed in column 8 of Table 2.
YSOc located closer to the center of the YSOc populated region in Multi-D color space will have higher YSO probability, and only isolated YSOc will have YSO probability equal to 1, which is the minimum value for YSO probability.   Out of 1313 identified YSOc, only 49 YSOc have YSO probability = 1 and 18 of them are newly identified YSOc. 
The chances for these ``isolated" YSOc in Multi-D color space not being real YSO are higher;
however, it is still possible that small numbers of YSOc contain unusual SED features.
Since the SEDs of YSOs are not exactly identical, lower YSO probability suggests that the type of SEDs are less common if they are indeed YSOs.  
Note that the YSO probability cannot be used to identify YSOc, because galaxies occupy similar color space.

\begin{deluxetable*}{cccccp{6.5cm}}
\tabletypesize{\tiny}
\tablecaption{YSOc excluded from the c2d lists}
\tablehead{
\colhead{number} & \colhead{R.A.}           & \colhead{Dec}      & \colhead{c2d classification\tablenotemark{a}} & \colhead{Type\tablenotemark{b}} &\colhead{reason description}
\\
\colhead{} &  \colhead{degree}           & \colhead{degree}      & \colhead{} & \colhead{} & \colhead{}
}
\startdata
\multicolumn{6}{c}{Perseus} \\
\hline
1 & 51.9117838 & 30.2160953 & YSO\_red &	B&	      Its MP1 flux is band-filled and its location is within the PSF of a ``Same'' YSOc (51.9128437, +30.2175414) at 6 arcsec away.\\
2& 52.2383644 & 31.2386438 & YSOc\_star+dust(MP1) &	B&	Its MP1 flux is band-filled and its location is within the PSF of a ``Same'' YSOc (52.2390096, 31.2377395) at 4 arcsec away.\\
3& 52.2578166 & 31.2814344 & YSOc\_star+dust(IR2) &	B&	MP1 locates in the diffraction spike of an extremely bright source and thus possibly has wrong flux measurment. \\
4& 52.2693214 & 31.3682561 & red1 &	C&	     This is not a HREL source. It is only detected in MP1 and the MP1 flux appears to be affected by the cloud emission.\\
5& 52.2957986 & 31.3072254 & YSOc\_red &	B&	     Its IR2, IR3, IR4, and MP1 flux are band-filled, and its location is between 
two ``Same'' YSOc (52.2944829, 31.3057251) and (52.2969172, 31.3087350). \\
6& 52.3229300 & 31.4634218 & YSOc\_red &	B&	    It appears like a jet knot 5 arcsec from a ``Same'' YSOc (52.3215278 31.4629087). \\
7& 53.1734652 & 31.1789357 & YSOc &	B&	     Its MP1 flux is band-filled and its location is within the PSF of a ``Same'' YSOc (51.9128437, 30.2175414) at 2 arcsec away.\\
8& 55.9791799 & 32.0175333 & YSOc\_star+dust(IR2) &         B&  This is a jet knot of HH211 close to the driven source. HH211 has only been detected at MP2 or longer wavelength, thus is not identified by our work. \\
9& 55.9854925 & 32.0146713 & red &	        B&	     This is a jet knot of HH211 close to the driven source. HH211 has only been detected at MP2 or longer wavelength, thus is not identified by our work. \\
10&55.9886063 & 32.0131886 & red &		B&	     This is a jet knot of HH211 close to the driven source. HH211 has only been detected at MP2 or longer wavelength, thus is not identified by our work. \\
11&55.9901535 & 32.0124525 & Galc &		B&     This is a jet knot of HH211 close to the driven source. HH211 has only been detected at MP2 or longer wavelength, thus is not identified by our work. \\
12&55.9909120 & 32.0534315 & red &		C&	     This is not a HREL source. Although this source has detections with S/N=3, 2 and $>$ 7 at IR2, IR3 and MP1, we think it is not a real source due to the contamination from cloud emission.\\
13& 55.9975507 & 32.0099032 & YSOc\_red &	B & This is a jet knot of HH211 close to the driven source. HH211 has only been detected at MP2 or longer wavelength, thus is not identified by our work. \\
14& 56.0797337 & 32.2883878 & YSOc &	A&	     Giant star in this work (\S 3.3). \\
15& 56.1498481 & 32.1567564 & YSOc\_star+dust(IR2) &	B&	IR4 and MP1 detections seem to be part of the clouds.  The source appears to be star-like at shorter wavelengths. \\
16&56.4473566 & 31.7198486 & YSOc\_star+dust(IR4) &	C&	This is not a HREL source. \\
\hline
\multicolumn{6}{c}{Lupus}\\
\hline
17&234.7014819 & -34.6772876 & YSOc\_PAH-em &	B&    This source is elongated in IR1 image, thus we identified it as a galaxy. \\
18&242.2427972 & -39.1265213 & YSOc &	A&	     Giant star in this work (\S 3.3). \\
19&242.3921400 & -39.2283528 & YSOc &	A&	     Giant star in this work (\S 3.3). \\
\hline
\multicolumn{6}{c}{Ophiuchus}\\
\hline
20&246.5609716 & -24.4187556 & red1 &	C&	    This is not a HREL source. It is only detected in MP1 and we think the MP1 flux is contributed from cloud.\\
21&246.7462569 & -24.5842515 & red &	B&	     Its MP1 flux is band-filled and its location is within the PSF of a newly identified YSOc (246.7464700, -24.5829609) at 6 arcsec away. \\
22&246.8052355 & -24.6926085 & YSOc &	A&	     Giant star in this work (Gaint star GY 232 identified by Luhman et al. 1999). \\
23&246.8747631 & -24.5601131 & YSOc &	A&	    Giant star in this work (\S 3.3). \\
24&246.9323892 & -24.7188227 & YSOc &	A&	     Giant star in this work (\S 3.3). 
\enddata
\end{deluxetable*}
\addtocounter{table}{-1}
\begin{deluxetable*}{cccccp{6.5cm}}
\tabletypesize{\scriptsize}
\tablecaption{YSOc excluded from the c2d lists}
\tablehead{
\colhead{number} & \colhead{R.A.}           & \colhead{Dec}      & \colhead{c2d classification\tablenotemark{a}} & \colhead{Type\tablenotemark{b}} &\colhead{reason description}
\\
\colhead{} &  \colhead{degree}           & \colhead{degree}      & \colhead{} & \colhead{} & \colhead{}
}
\startdata
\multicolumn{6}{c}{Serpens}\\
\hline
25&277.0458600 & -0.0276004 & YSOc\_star+dust(MP1) &	A& Giant star identified by Oliveira et al. (2009) \\
26&277.1141023 & -0.1972144 & YSOc\_star+dust(MP1) &	A& Giant star identified by Oliveira et al. (2009) \\
27&277.2270870 & 0.4812201 & YSOc &	A&	     Giant star in this work (\S 3.3). \\
28&277.2837039 & -0.1269813 & YSOc\_star+dust(MP1) &	A&	Giant star identified by Oliveira et al. (2009) \\
29&277.2876881 & 0.5244498 & Galc &	B&     This is a jet knot of ``Same'' source (277.2877868, 0.5256632) close to the driven source.\\
30&277.3855879 & -0.2231446 & YSOc\_star+dust(MP1) &	A& Giant star identified by Oliveira et al. (2009) \\
31&277.5219079 & 0.6846068 & red &		E&	    This source locates in the edge of the mapping area of c2d, thus not observed in IR1 and IR3.  We do not consider such sources in our analysis.  \\
32&277.5237366 & 0.6588334 & rising &	E&	     This source locates in the edge of the mapping area of c2d, thus not observed in IR1 and IR3.  We do not consider such sources in our analysis.  \\
33&277.5451572 & 0.7835665 & YSOc\_star+dust(IR4) &	D&	We identify this source as a galaxy due to its galaxy probability (Gal prob1 = 1.07 and Gal prob2 =1.04).
\enddata
\tablenotetext{a}{
If the c2d source type does not contains ``YSO'', the source is added to the YSOc list by \citet{ev09} based on ancillary data from literature.}
\tablenotetext{b}{Reasons for excluding the source from our YSOc list.\\
A: This source is classified as a giant star by either this work or Oliveria et al. (2009).\\
B: This source is excluded in image checking process.\\
C: This source is not considered in our identification process. It has detections at fewer than three bands or is not a HREL source (see \S2.1).\\
D: This source has both galaxy probabilities, Gal prob 1 and Gal prob 2, larger than 1.\\
E: This source is not observed at IR1 and IR3 because it is in the edge of c2d map\\}
\end{deluxetable*}
\subsection{Missed or Excluded YSOc from The c2d Lists}
In this section, we discuss the c2d identified YSOc but missed through our selection processes.
There are 36 sources missed in total. We recover only three of them back to our YSOc candidate list (Table 2) because they are previously identified well-known sources.
These three sources are not identified in our identification process because they are not HREL sources (Table 4); DK Cha is saturated at several Spitzer bands, and VLA1623 and IRAS 16293 are only detected at MP1. For the remaining 33 sources, they are separated into two groups:  23 sources are identified by \citet{ha07} using several CMD criteria (``c2d classification'' starting with ``YSOc'' in column 4 of Table 5), and 10 sources are added by \citet{ev09} with additional observations from other literatures \citep{du08,en09,jo07,jo08}.  Here we justify why they are not likely to be true YSOs, and the exact reasons for the removal of each source are listed in Table 5.

First, for the 23 YSOc identified by \citet{ha07},
\begin{enumerate}[leftmargin=0.4cm]
\item Eleven sources may be giant stars: seven sources are classified as giant stars by this work, including a giant star GY232 independently  identified by \citet{lu99} with spectroscopic data, and four sources are identified as  giants by \citet{ol09}.
\item Ten sources are removed by image checking process:  seven sources have flux contamination from nearby sources or cloud emission, one source is elongated like a galaxy, and two sources are jet knots from HH211.
\item One source is not a HREL object. It is also very distant from the cluster regions with low stellar surface density in Perseus and slightly extend, thus we suggest it is a background galaxy.
\item One source has Gal Prob $>$1 (1.07).
\end{enumerate}

Second, for the ten c2d YSOc added by \citet{ev09} but removed by our work,
\begin{enumerate}[leftmargin=0.4cm]
\item Three sources were identified as jet knots in HH211 system by image checking.  
Note that HH211 itself also cannot be identified through our procedure, since it has only been detected at MP2 and longer wavelengths. 
\item Two sources locate very close to selected YSOc, and their flux appear to be contaminated by the nearby YSOc. 
\item Five sources have not been considered in our identification process.  Three of them are not HREL sources, and they all appear to be extended and seems to be part of the cloud structure.  The remaining two are located at the edge of survey regions without IR1 and IR3 observations, which are not considered in our analysis. 
\end{enumerate}

\subsection{PAH Emission}
From SWIRE data, we find that the galaxies with a peak at IR4 in SEDs (PAH emission) are usually bright, thus are much difficult to be eliminated.
The PAH emission are usually seen in star-forming galaxies \citep{ev09}, thus the c2d project label such sources with ``PAH\_em'' (col. 9 in Table 2) using the color criteria [3.6]-[4.5] $>$ 1.5 and [5.8]-[8.0] $<$ 0.6 \citep{ev07}.  There are 37 sources with ``PAH\_em'' label out of our 1313 YSOc (30 from newly identified YSOc and 7 are ``Same'' YSOc).   The YSOc satisfying the PAH-em conditions are not excluded in neither the c2d YSOc list nor our YSOc list, because they are not clearly identified as galaxies.  More observations toward these sources are necessary for identifying their nature, and we should use these sources with caution.  Although those YSOc could be wrongly identified, the small fraction of 2.8\% should not significantly influence our statistical studies. 

\subsection{Uncertainty From The Selections of Bands for The Two 5-D arrays}
Because we use two specific 5-D arrays to identify YSOc instead of a 10-D array, it can result in an uncertainty in our YSOc sample from the specific selection of bands.
To understand the effect, 
we construct another three Multi-D arrays to select YSOc and compare the results with our YSOc identified using bands of J, K, IR2, IR4 and MP1 (1277 sources). These three Multi-D arrays are constructed using bands of A=(J, H, IR1, IR3 and MP1), B=(J, Ks, IR1, IR4 and MP1) and C=(J, IR1, IR2, IR4 and MP1), respectively.  They all include J and MP1 bands to represent the SED structures with the longest coverage of wavelengths and subarrays are also constructed for each main array.
Following the whole selection process, there are 7, 7 and 2 additional sources selected using A, B and C arrays, respectively.
One source is selected by all three spaces and another one is selected by both A and B spaces; thus, totally 13 new sources are selected using these three spaces. 
These 13 sources all have galaxy probability greater than 0.4 which is corresponded to $\sim$ 2 cells (0.4 magnitude) from a boundary of galaxy populated region with our Gaussian beam.
Therefore, we suggest that the specific band selection could result in $\lesssim$ 1\% difference compared to our original YSOc selection and the difference only occurs in the region near the boundaries between YSOc and galaxies.

\section{DISCUSSION}
\subsection{New YSO Identification Method and Results}
We develop Multi-D method for separating two kinds of sources which have similar SEDs and are indistinguishable in CMDs and CCDs.
In this paper, we use Multi-D method to identify YSOc in molecular clouds with Spitzer data by comparing to galaxy sample from SWIRE.  
A large number of new YSOc are identified and our analysis indicates that they are much like faint YSOs not selected before (Fig. 10). This result suggests that Multi-D method is very powerful in identifying faint YSOs that have been detected. However, there are several caveats in Multi-D method.
\begin{enumerate}[leftmargin=0.4cm]
\item The SWIRE data may not represent a complete collection of galaxy SEDs.  In addition, in the direction of the observed clouds, the brightest galaxies may be brighter than all SWIRE galaxies due to their proximity.  
Thus, if SWIRE data are incomplete galaxy samples, 
some true galaxies will not be removed and will be identified as YSOc.
\item Some YSOs may have similar SEDs and brightness to those of galaxies. 
If a YSO has SED and brightness similar to galaxies in the comparison galaxy sample in the analysis wavelengths, it will have a galaxy probability larger than one and will be removed from YSOc list. We found that our YSOc and galaxy sample are continuously distributed in MMDs and 3-D plotting (Fig. 4--9), which implies that there may be faint YSOs located in the galaxy populated region and
can not be identified with Spitzer data only.  However, no available methods can identify such sources with only photometry data.
\item Some sources may have high uncertainty in photometry which will affect the YSO identification process. 
Sources with wrong flux measurements will have peculiar SED structures, which makes them locate far from the galaxy populated region in Multi-D space and thus be classified as YSOc. We have designed an image checking process to reduce such cases (\S 3.4).
\end{enumerate}
Although Multi-D method has these caveats, such problems are unavoidable and also exist in those methods using CCD and CMD criteria.
Multi-D method does allow us to obtain the most complete YSOc list from provided data set.

\begin{deluxetable*}{cccccc}
\tabletypesize{\scriptsize}
\tablecaption{VeLLO candidates}
\tablehead{
\colhead{R.A.}           & \colhead{Dec}       &\colhead{L$_{IR}$} & \colhead{L$_\textmd{int}^{70\mu m}$} & \colhead{c2d classification} & \colhead{A$_\textmd{V}$}
\\
\colhead{degree}           & \colhead{degree}      & \colhead{L$_\odot$} &\colhead{L$_\odot$} & \colhead{} & \colhead{mag}
}
\startdata
52.2752091  & 30.5108863 & 0.017 (0.007) & 0.023 (0.006) & YSOc\_star+dust(IR2) & 2.4 \\
52.3348374  & 31.2139919 & 0.016 (0.006)& 0.034 (0.008)& --	&	9.5	\\
52.7512777  & 30.9369012 & 0.009 (0.004)& 0.011 (0.003)& --	&	0.9	\\
53.2410000  & 31.1022931 & 0.010 (0.004)& 0.016 (0.004)& YSOc\_red	&	13.7\\
56.3075775  & 32.2027786 & 0.050 (0.02)& 0.054 (0.013) & YSOc\_star+dust(IR2)	 & 3.0\\
194.2566223 & -76.8097467& 0.028 (0.007)& 0.049 (0.008)& --	 &	2.4	\\
277.2287724 & 0.3090777  & 0.034 (0.003)& 0.038 (0.005)& YSOc\_red	&	15.9
\enddata
\end{deluxetable*}

\subsection{Newly Identified Very Low Luminosity Object (VeLLO) Candidates}

VeLLOs are faint embedded protostars (L$_\textmd{int}$ $<$ 0.1 L$_\odot$), therefore they are difficult to be identified.   Recent works suggest that VeLLOs are likely to be protostars at quiescent accreting phase
\citep{du10}.  However, this does not exclude the possibility that some VeLLOs could be faint protostars 
at early evolutionary stages or even ``first cores" which is the transition phase between
starless cores and Class 0 sources.   Since our method discovers more faint protostars
compared to previous works, we here examine whether we can identify more VeLLOs.

The most thorough survey of VeLLOs is carried out by \citet{du08} using c2d data.
Several other sources are claimed to be VeLLOs in the recent years \citep{pi11,ka11}.
Since a large number of faint YSOc are identified using Multi-D method in this work,
we are in good position to add reliable VeLLO candidates to the collection. 
Fig.\ 10b shows that some newly identified Class 0 and Class I YSOc (with MP2 detections) have lower
infrared luminosity (luminosity between 1.25--70$\mu$m) than 
that of the more reliable VeLLO candidates (group 1--3) identified by \citet{du08}. 
\citet{du08} use a relation to translate the MP2 flux to the internal luminosity L$_\textmd{int}$ = 3.3 $\times$ 10$^8$ F$^{0.94}_\textmd{70}$ L$_\odot$ where F$_{70}$ is MP2 flux in cgs units (ergs cm$^{-2}$ s$^{-1}$), based on the SED model from Monte Carlo dust radiative transfer code RADMC \citep{du04}.
We adopt this relation to estimate L$_\textmd{int}$ for our Class 0 and I candidates and find 32 sources with L$_\textmd{int}$ $<$ 0.1 L$_\odot$ in addition to VeLLOs found in \citet{du08}.
In order to make sure the selected VeLLOs are young YSOs, two more criteria from \citet{du08} are adopted 
to ensure that the SEDs rise (1) from the longest detected IRAC wavelength to MP1 and (2) from MP1 to MP2, which 
reduce the number of VeLLO candidates to seven (Table 6). 

\begin{figure}
\includegraphics[scale=0.55]{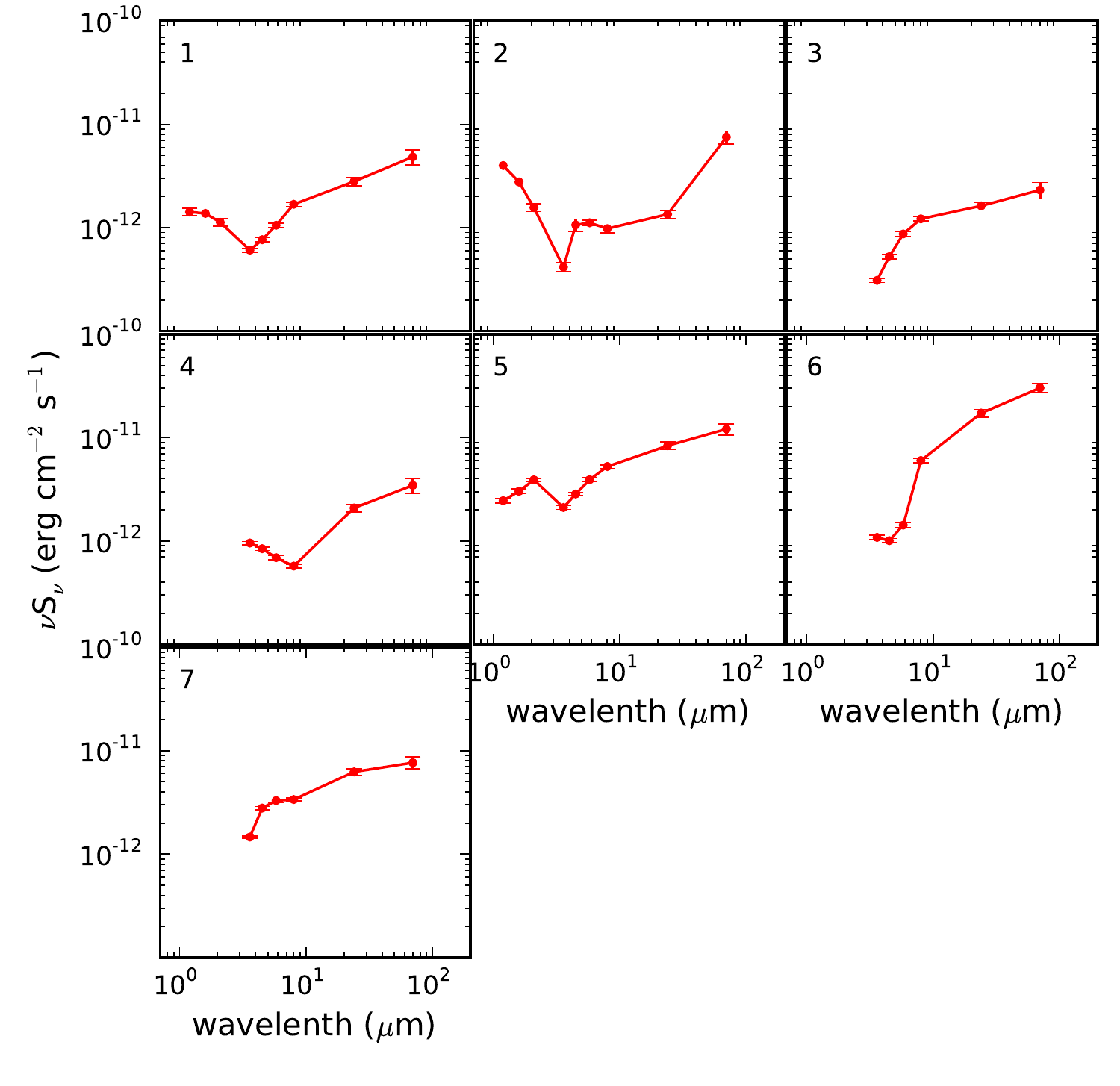}
\caption{The SEDs of VeLLO candidates identified from our YSOc list (see \S 5.2 for selection criteria). Number 3 is newly identified from our work, and the rest of VeLLOs have been identified by Dunham et al. (2008) but was classified as less confident VeLLOs.}
\end{figure}

Their SEDs are shown in Fig.\ 13.
Among these seven candidates, only one is newly identified by us. 
The other six cadidates have been identified by \citet{du08},
but they are in less confident group 4-7 corresponding to sources not obviously associated with high column density regions.
Since one possible nature of VeLLOs is proto brown dwarfs which could be ejected from their parent cores during formation processes,
VeLLOs may not necessarily be embedded in high density regions.
Therefore, we do not exclude these sources from our VeLLO candidate list.   In short, with Multi-D method, faint protostar candidates can be selected naturally. 

\subsection{Statistical Properties}
The significant increase of YSOc numbers in molecular clouds can alter statistical properties, such as SFR and lifetimes of YSOs in different evolutionary stages. The variations of these two properties are analyzed in this section.

\begin{deluxetable*}{cccccccc}
\tabletypesize{\scriptsize}
\tablecaption{Facts and parameter about clouds}
\tablehead{
\colhead{Cloud}	 & \colhead{distance\tablenotemark{a}} &\colhead{N(YSO)\tablenotemark{b}} &
\colhead{M(cloud)\tablenotemark{a}} & \colhead{Area\tablenotemark{c}} &
\colhead{$\Delta v$\tablenotemark{a}} &
\colhead{SFR$_\textmd{ff}$} &
\colhead{$\alpha_{vir}$} 
\\
\colhead{}                & \colhead{pc} & \colhead{\#} &
\colhead{M$\odot$} & \colhead{pc$^2$} &
\colhead{km s$^{-1}$} &
\colhead{} &
\colhead{} 
}
\startdata
Perseus			&250 (50) &452	&4814 (1925)	&62.0 (24.8)		& 1.54 (0.11)		& 0.050 (0.034)	& 0.46 (0.21) \\
Serpens			&260 (50)\tablenotemark{d} &295	&2016 (775)		&17.5 (6.7)\tablenotemark{e}	& 2.16 (0.01)		& 0.047 (0.030)	& 1.14 (0.49)  \\
Ophiuchus		&125 (25)	 &343	&2182 (873)		&27.3 (10.9)		& 0.94 (0.11)		& 0.068 (0.045)	& 0.25 (0.13) \\
Lupus I			&150 (20) &20	&250 (58)			&4.5 (1.2)			& 1.9 (0.19)\tablenotemark{f}	& 0.026 (0.012)	& 3.62 (1.30) \\
Lupus III			&200 (20) &60	&443 (102)		& 8.5 (1.7)		& 1.7 (0.17)\tablenotemark{f}	& 0.054 (0.018)	& 2.25 (0.68) \\
Lupus IV			&150 (20) &13	&119 (28)			& 1.7 (0.5)		& 1.7 (0.17)\tablenotemark{f}	& 0.025 (0.011)	& 3.74 (1.34) \\
Chamaeleon II	&178 (18) &27	&426 (86)			&7.4 (1.5)			& 1.2 (0.4)\tablenotemark{g}		& 0.023 (0.008)	& 1.09 (0.77)
\enddata
\tablenotetext{a}{The area and $\Delta v$ of clouds are from \citet{ev09}.}
\tablenotetext{b}{The YSOc numbers are counted with YSOc located in region with A$_\textmd{V}$ $>$ 2 magnitude.}
\tablenotetext{c}{The area is calculated from extinction map with A$_\textmd{V}$ $>$ 2 magnitude area.}
\tablenotetext{d}{We use the distance uncertainty of Serpens with a wide range to cover the distance measurement mention in \citet{ev09}, i.e., \citet{st96, ei08}}
\tablenotetext{e}{The extinction map of Serpens has all pixel with A$_\textmd{V}$ $>$ 2 magnitude, thus the area is from Spitzer observing region.}
\tablenotetext{f}{The $\Delta v$ in Lupus is from \citet{ta96}}
\tablenotetext{g}{The $\Delta v$ in Chamaeleon is from \citet{vi94}}
\end{deluxetable*}

\subsubsection{Comparing Star Forming Rate with Theoretical Models}
The accuracy of SFR is critical in distinguishing whether the star formation in GMCs is dominated by turbulence or magnetic fields.  
Without any supports, typical GMCs should collapse on its free-fall time scale, which results in a SFR of roughly 250 M$_\odot$ yr$^{-1}$\citep{kr05}.  
However, the SFR in Milky Way is measured as $\sim$ 3 M$_{\odot}$ yr$^{-1}$ \citep{mc97}.
Thus, supporting forces, such as turbulence and/or magnetic fields, are required to reconcile the difference between theory and observations. 
To study how the turbulence affects the SFR, \citet{kr05} defines the dimensionless SFR per unit free-fall time, SFR$_\textmd{ff}$,
\begin{equation}
 \textmd{SFR}_\textmd{ff} =\frac{\textmd{SFR}\times t_\textmd{ff}}{\textmd{M(cloud)}}~,
\end{equation}
where t$_\textmd{ff}$ is the free fall time of the cloud and M(cloud) is the mass of the cloud, and derive an analytic expression for SFR$_\textmd{ff}$ in a supersonic turbulent median,
\begin{equation}\label{}
\textmd{SFR}_\textmd{ff} \approx 0.014 (\frac{\alpha_{vir}}{1.3})^{-0.68}(\frac{\mathcal{M}}{100})^{-0.32},
\end{equation}
where $\alpha_{vir}$ is the virial parameter defined by \citet{be92} measuring the ratio of kinetic energy and gravitational energy of a clumpy, and $\mathcal{M}$ is the Mach number.
The definition of $\alpha_{vir}$ is $\alpha_{vir}$=5$\sigma _{tot}^{2}$R/(GM) where $\sigma _{tot}$ is velocity dispersion from thermal and turbulent velocities over entire cloud, R and M are radius and mass of the cloud, respectively.
The turbulence dominated model suggests that a small $\alpha_{vir}$ can result in a large $\textmd{SFR}_\textmd{ff}$.
However, magnetic field dominated models  predict that the SFR$_\textmd{ff}$ is proportional to $\alpha_{vir}$ \citep{kr07}, which is
\begin{equation}\label{}
\textmd{SFR}_\textmd{ff} \approx 0.01 \alpha_{vir}.
\end{equation}
The parameter values \citet{kr07} used to derive this relation could vary more than an order of magnitude, resulting in a wide range of possible value for the coefficient. Nevertheless, the linear proportionality of SFR$_\textmd{ff}$ and $\alpha_{vir}$ should be unaffected.
Hence, comparing the relation of SFR$_\textmd{ff}$ and $\alpha_{vir}$ may provide us the hint that which mechanism dominates in molecular clouds.

\begin{figure}
\includegraphics[scale=.55]{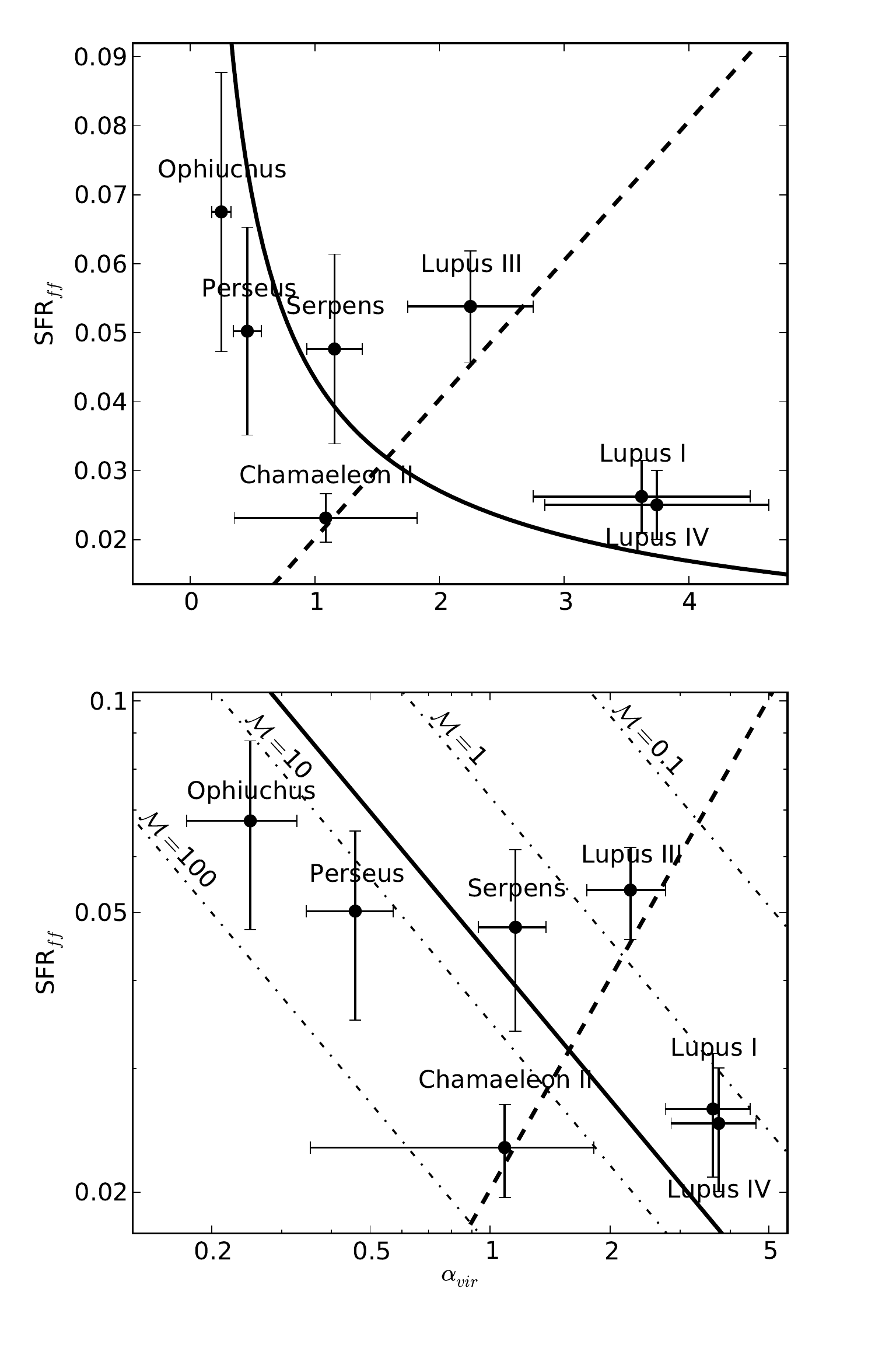}
\caption{The plot of $\alpha_{vir}$ to SFR$_\textmd{ff}$. The solid line and dashed line represents the best fitting curve of eq. 4 (turbulence model) and eq. 5 (magnetic field model), respectively.}
\end{figure}

The c2d data and our new YSOc catalogs provide necessary information to compare SFR$_\textmd{ff}$ and $\alpha_{vir}$. 
SFR$_\textmd{ff}$ is estimated following the same analysis method in \citet{ev09}.
Assuming that the mean mass of a YSO M$_{*}$ $=$ 0.5 M$_\odot$ and the period of star formation T $=$ 2Myr,
the star formation rate can be estimated from 
\begin{equation}\label{}
\textmd{SFR}=\frac{\textmd{N(YSO)}\times \textmd{M}_{\ast}}{\textmd{T}} 
\end{equation}
and
\begin{equation}\label{}
 t_\textmd{ff} = 34 \textmd{Myr}/ \sqrt{\textmd{n}},
\end{equation}
where N(YSO), M$_{\ast}$, T and n are the number of YSO, mean mass of YSO, period of star formation and number density of cloud, respectively.
Inserting Equation (6) and (7) into Equation (3), we are able to obtain the SFR$_\textmd{ff}$ from N(YSO), M(cloud) and n.
We calculated n from the cloud mass and surface area by assuming a spherical cloud and the cloud mass was obtained from Table 1 in \citet{ev09}. The cloud mass and the surface area of the clouds are both obtained from the c2d extinction maps in area with A$_\textmd{V}$ $>$ 2.
We use the mass of the clouds from the table 1 in \citet{ev09} and calculated the surface area of the clouds, which are both from the c2d extinction maps in area with A$_\textmd{V}$ $>$ 2 region. 
The errors of SFR$_\textmd{ff}$ are from error propagating and only uncertainty considered here is the uncertainty in cloud distances.


We calculate $\alpha_{vir}$ using $\alpha_{vir}$=5$\sigma _{tot}^{2}$R/(GM). R and M are the same to that used in SFR$_\textmd{ff}$ calculations. The $\sigma _{tot}$ are velocity dispersions from $^{13}$CO J = 1 $\to$ 0 observations and are averaged over a full map of the cloud. For Perseus, Serpens and Ophiucuhus, 
we use the $\sigma_{tot}$ from \citet{ev09} which obtained the values from the ÒThe COordinated Molecular Probe Line Extinction Thermal Emission Survey of Star Forming RegionsÓ (COMPLETE) project \citep{ri06}. 
For Lupus and Chamaeleon, we use the $\sigma _{tot}$ from \citet{ta96} and \citet{vi94}, respectively, which also use $^{13}$CO J = 1 $\to$ 0 as the tracer. The calculated $\alpha_{vir}$ for all clouds are shown in Table.\ 7 and the propagating errors are from the errors of distance and $\sigma_{tot}$ in the cloud.

Fig.\ 14 shows the relations between $\alpha_{vir}$ and SFR$_\textmd{ff}$ for all clouds in this paper. 
We fit the data with the predictions for the turbulence model and the magnetic field model, i.e., Equation (4) and (5), respectively, derived by \citet{kr07}. For the magnetic field model, we set the coefficient to be a variable in Equation (5).
The reduced $\chi^{2}$ are 3.0 and 5.6 for the turbulence model and the magnetic field model, respectively, which implies that the data are more consistent with the turbulence model than the magnetic field model. 
Our result hints that turbulence dominates the star formation in large scales such as GMCs.
In addition, in Fig.\ 14, the SFR$_\textmd{ff}$ of Lupus are all larger than that of the best fitting curve of the turbulence model, especially for Lupus III.   This could result from our assumption of a mean mass 0.5 M$_\odot$ for YSOc in all clouds, but the mean stellar mass in Lupus may be closer to 0.2 M$_\odot$ \citep{me08}. The mean stellar mass is 0.6 or 0.8 M$_\odot$ for different evolutionary tracks in Serpens \citep{ol09} and is 0.52 $\pm$ 0.11 M$_\odot$ in Chamaeleon II \citep{sp08}.  Better estimates of the mean stellar mass in all clouds will provide a more accurate SFR$_\textmd{ff}$ and give a stronger conclusion.

\begin{deluxetable*}{cccccccccc} 
\tablecolumns{10} 
\tabletypesize{\scriptsize}
\tablecaption{Numbers of YSOc by clouds and class} 
\tablehead{ 
\colhead{}    &  \multicolumn{4}{c}{This work} &   \colhead{}   & 
\multicolumn{4}{c}{\citep{ev09}} \\ 
\cline{2-5} \cline{7-10}
\colhead{Cloud} & \colhead{I/0}   & \colhead{Flat}    & \colhead{II} & 
\colhead{III} & \colhead{}   & \colhead{I/0}   & \colhead{Flat}    & \colhead{II} & \colhead{III}
}
\startdata 
Perseus			&99	&49	&272	&49	&	&87	&42	&225	&31\\
Serpens			&55	&36	&170	&35	&	&36	&23	&140	&28\\
Ophiuchus		&72	&60	&191	&58	&	&35	&47	&176	&34\\
Lupus				&12	&12	&64	&41	&	&5		&10	&52	&27\\
Chamaeleon II	&7		&3		&21	&7		&	&2		&1		&19	&4\\
Total				&245 (19\%)	&160 (12\%)	&718 (55\%)	&190 (14\%)	&	&165 (16\%)	&123 (12\%)	&612 (60\%)	&124 (12\%)
\enddata
\end{deluxetable*} 

\subsubsection{Lifetime in Different Evolutionary Stages}
The lifetimes of YSOs in different evolutionary stages are estimated from the numbers of YSOs in each stages, and thus the variation of YSOc sample may result in different lifetime estimations. 
YSOs are commonly classified into four evolutionary stages Class 0/I, Flat, Class II and Class III from young to old using the spectral slope, $\alpha$, which is the best fit slope from K to MP1 band of the SED.
The definition of $\alpha$ and the classification criteria are from \citet{gr94}:
\begin{equation}\label{}
\alpha=\frac{d~\textmd{log(}\lambda\textmd{S}(\lambda))}{d~\textmd{log}(\lambda)}
\end{equation}
and\\
Class 0/I: 0.3 $\leq \alpha$ \\
Flat: -0.3 $\leq \alpha <$ 0.3 \\
Class II: -1.6 $\leq \alpha <$ -0.3 \\
Class III: $\alpha <$ -1.6 \\

Although it has been demonstrated that the SED morphologies of YSOs are affected by geometry such as inclination angle with radiative transfer codes \citep{wh03a,wh03b,ro06,cr08}, here we still 
use $\alpha$ as a rough age indicator as there is no better alternatives. 
Our large YSO sample size may mitigate the geometry effect.

Fig.\ 15 shows the distributions of $\alpha$ in each clouds and the fractions of YSOc in different evolutionary stages,
and the latter numbers are almost the same to the results in \citet{ev09} in either individual cloud
or the whole data set, implying that the estimates of lifetimes of each evolutionary stages are not changed (Table 8).
Fig. 15 also shows that the distributions of $\alpha$ in different clouds are different, which indicates that those clouds 
may be in different evolutionary stages.  Because Perseus molecular cloud consist of two major YSO clusters NGC1333 and IC348, we separated it into two regions, western and eastern Perseus, by R.A.=54.3$^{\circ}$.    
Our results suggest that western Perseus (NGC1333) has the highest fraction of Class 0/I YSOc, while Lupus has the highest fraction of Class III YSOc.  Therefore, western Perseus is the youngest cloud and Lupus is the oldest cloud among c2d-surveyed regions, while the other clouds are at similar evolutionary stages.\\

\begin{figure}
\includegraphics[scale=.38]{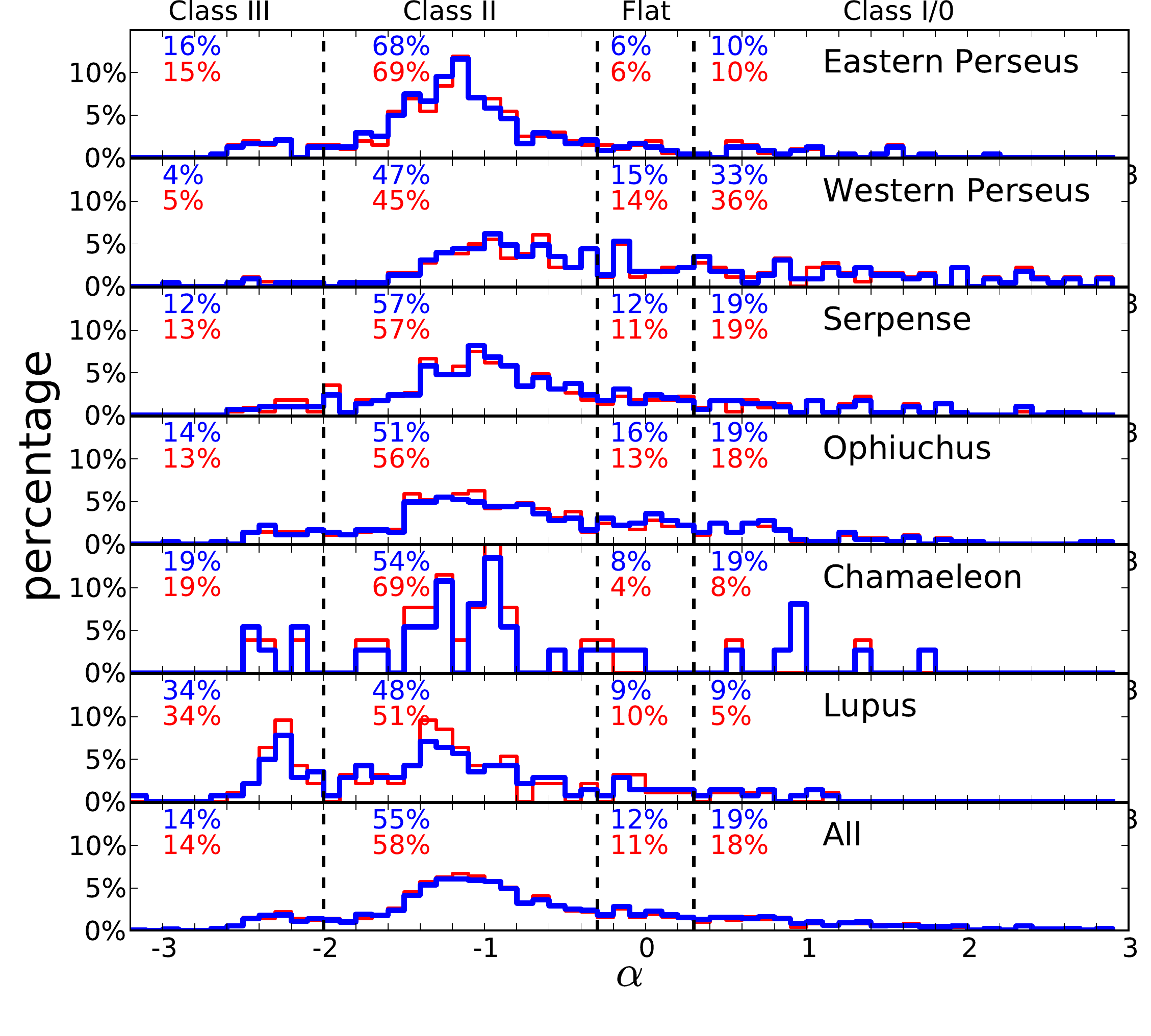}
\caption{The populations of all sources with age indicator, $\alpha$, in the five clouds. 
Blue and red lines indicate our YSOc and the ``Same'' YSOc, respectively.}
\end{figure}

\section{SUMMARY}
We have developed ``Multi-D method'' to identify YSOc from star-forming regions with reliable photometry measurements from multiple bands.  Main-sequence and Giant stars are first eliminated from the observed sources, and the rest of the sources are compared to a galaxy sample, such as SpizerÕs SWIRE dataset, in multi-dimensional magnitude space.  We demonstrate that this Multi-D method which uses multi-band photometry simultaneously can identify all possible YSOc above the galaxy confusion limit and recover those could be missed by using certain set of color-magnitude or color-color criteria.
We identify 1313 YSOc from Spitzer's c2d high reliability (HREL) catalogs, which is 28\% more than what have been identified by the c2d project \citep{ev09}.
The increase amount of YSOc suggests the following results:

\begin{enumerate}[leftmargin=0.4cm]
\item  The increase amount of YSOc directly increases the SFR estimated in each region by the same percentage.
We further compare the relation between $\alpha_{vir}$ and SFR$_\textmd{ff}$ to theoretical models, and our results are more consistent with the prediction of supersonic turbulence dominated model than that of magnetic field dominated model.
\item Although we identify 28\% more YSOc than those listed in the c2d catalogs, the lifetimes of YSOs in different evolutionary stages are unchanged
since the fractions of YSOc in different Classes are almost the same.
\item Our Multi-D method allows us to reliably identify more faint YSOc, which can be used to select VeLLOs.
Using the 70$\mu$m flux to the internal luminosity relation suggested by \citet{du08}, we are able to find 7 new VeLLO candidates. 

\end{enumerate}

\end{document}